\documentclass[12pt]{jpconf}

\usepackage{physics}
\usepackage{verbatim}
\usepackage{graphicx}
\usepackage{caption}
\usepackage{amsmath}
\usepackage{mhchem}
\usepackage{siunitx}
\usepackage{hyperref}
\usepackage{dsfont}
\usepackage{caption}
\usepackage{subcaption}
\usepackage{mathrsfs}
\usepackage{amssymb}
\usepackage{bbm}
\usepackage{mathtools}
\usepackage{tikz}
\usepackage{mciteplus}
\usepackage{pstricks}
\usepackage{color}

\usepackage{collref}
\usepackage{cite}

\usepackage{cancel}
\usepackage{xcolor}
\hypersetup{
    colorlinks,
    linkcolor={red!50!black},
    citecolor={blue!50!black},
    urlcolor={blue!80!black}
}

\newcommand{\diag}{\mathop{\mathrm{diag}}}

\begin{document}

\title{Scherk-Schwarz orbifolds at the LHC}

\author{ D~D~Smaranda and D~J~Miller }

\address{SUPA, School of Physics and Astronomy, University of Glasgow, Glasgow, G12 8QQ, \\ United Kingdom \\[3mm]
E-mail: d.smaranda.1@research.gla.ac.uk, david.j.miller@glasgow.ac.uk}

\begin{abstract}
We examine orbifold theories of Grand Unification with Scherk-Schwarz twisting, performing a renormalisation group analysis and applying low energy experimental constraints. We rule out the minimal SU(5) models, and consider simple extensions including additional fields, such as an additional scalar field, or additional symmetries, such as $SU(5)\times U(1)$ or $E_6$. We find that it is very difficult to generate a large enough Higgs mass while simultaneously passing LHC experimental search constraints.
\end{abstract}

\section{Introduction}

The Large Hadron Collider's (LHC) triumph on the discovery of the $125\,$GeV Higgs boson~\cite{HiggsATLAS, HiggsCMS} has been tempered somewhat by the lack of evidence of physics beyond the Standard Model (SM). Indeed, Supersymmetry, which was for many years the most popular beyond the SM speculation, is now facing significant exclusions from the LHC~\cite{Aad:2014gfa, Aad:2015rba, Aad:2015iea, Aad:2015jqa, Aad:2015eda, Aaboud:2016dgf,  Aaboud:2016hmk, Aaboud:2016uth, Aaboud:2017vwy, Aaboud:2017phn, Aaboud:2017dmy, Aaboud:2018ujj, Aaboud:2017hdf, Aaboud:2017hrg, Aaboud:2017wqg, Aaboud:2017nfd, Aaboud:2017aeu, Aaboud:2017ayj, Aaboud:2017ejf, Aaboud:2017mpt, Aaboud:2017iio, Aaboud:2017leg, Aaboud:2017nhr, Aaboud:2017nmi, Aaboud:2017opj, Sirunyan:2017pjw,Aaboud:2017sjh, Aaboud:2018zjf, Aaboud:2018sua, Aaboud:2018jiw, Aaboud:2018htj, Aaboud:2018zeb,  Aaboud:2018lpl}. While many of these exclusions are for so-called ``simplified models'' (see e.g.~Refs.~\cite{Aad:2015iea, Csaki:1996ks}), which make assumptions about the superpartner mass spectrum, these exclusions are particularly hard on models where the supersymmetry breaking parameters are assigned restricted values at some high supersymmetry breaking scale. Indeed the constrained Minimal Supersymmetric Standard Model, which gives the supersymmetry breaking parameters common values at a high scale, is now mostly ruled out~\cite{Han:2016gvr, Bechtle:2015nua}. However, typical unconstrained supersymmetric models are complicated by having over 100 additional free parameters (due to the lack of knowledge of the supersymmetry breaking mechanism), so plenty parameter space remains for more non-minimal models of supersymmetry and the LHC will of course continue to search for them.

This naturally leads to two complementary approaches to the search for supersymmetry. Firstly, not all the supersymmetry breaking parameters are important for LHC searches, so one may make very reasonable assumptions about the model (such as no new source of CP-violation, no Flavour Changing Neutral Currents, and first and second generation universality) to reduce the number of parameters. For example, the {\em phenomenological} MSSM (pMSSM)~\cite{Djouadi:1998di} has ``only'' 19 additional parameters, making its investigation at the LHC much more practicable. Alternatively, one may posit mechanisms of supersymmetry breaking at the high scale to predict relations between the supersymmetry breaking parameters, and the consequent low energy spectrum that can be confronted with data.

This is often married with a Grand Unified Theory (GUT) in which the SM gauge groups are unified into a larger group.
Such theories address longstanding shortcomings of the SM, explaining the origin of its semi-simple group, the particular quark and lepton multiplet structure, and the lepton and quark charges. The combination with supersymmetry in turn allows a solution of the hierarchy problem, while helping with gauge coupling unification. Indeed, these issues make it difficult to build a convincing GUT without including supersymmetry.

Supersymmetry itself must be broken at low energies. One interesting idea is that supersymmetry may be broken by the compactification of extra dimensions~\cite{Scherk1, Scherk2}. An additional extra dimension may have escaped previous detection by being {\em rolled-up} with a radius $R$ that is smaller than the resolution of our colliders. Compactifying the extra dimension by imposing additional symmetries results in heavy Kaluza-Klein states~\cite{Kaluza:1921tu,Klein:1926tv} and can break supersymmetry as well as any underlying symmetry of grand unification.

Although the introduction of supersymmetry is helpful, it does not provide a solution for the doublet-triplet-splitting problem. We still must explain why the representation of the larger gauge group that contains the Higgs bosons is split in mass, with the SM Higgs doublet at the electroweak scale while the other states (e.g.\ a colour triplet in an $SU(5)$ GUT) remain at the GUT scale; and of course, this must be done while keeping proton decay suppressed \cite{ProtDecay1, ProtDecay2}. While there are proposals to solve this problem in $4$ dimensions \cite{DTSol1, DTSol2, DTSol3, DTSol4, DTSol5, DTSol6, DTSol7, DTSol8, DTSol9, DTSol10, DTSol11, DTSol12, DTSol13, DTSol14, DTSol15, DTSol16, DTSol17} these generally require a rather complicated Higgs sector. Instead one may again use the compactified extra dimensions, making appropriate quantum number choices to split the multiplet.

The symmetries imposed on the compactified dimension and the representation of the states under these symmetries define the model. In ``ordinary'' compactifations, the states transform trivially under these extra symmetries, e.g.\ $\phi(x^\mu,y+ 2 \pi R) = \phi(x^\mu,y)$ where $y$ is the extra dimension. However, if the transformation becomes non-trivial, e.g.\ $\phi(x^\mu,y+ 2 \pi R) = T \phi(x^\mu,y)$ where $T$ is a model dependent operator, the theory is said to have a Scherk-Schwarz (SS) twist ~\cite{Scherk1,Scherk2}. In this study we will seek to investigate some simple Scherk-Schwarz models of compactification to test if they are phenomenologically compatible with electroweak symmetry breaking and low energy experimental constraints.

Ultimately we will ask if these SS models, and in particular Barbieri, Hall and Nomura's original minimal model \cite{Barbieri1}, can provide a realistic and ``natural'' spectrum of broken supersymmetry at the TeV scale that may be discovered at the LHC.

We note that orbifold GUTs are effective theories requiring explicit cut-offs due to the loss of perturbative control associated with the Remormalisation Group Equation (RGE) running of the couplings. Consequently, these theories may potentially be embedded into fully complete UV string theories \cite{DIXON1986285, DIXON1985678} where the cutoff is replaced by the physical string scale. Previous attempts have implemented phenomenologically acceptable models using D-brane and F-brane models \cite{Maharana:2012tu}, with the focus of models presented here on $E_8 \times E_8$ heterotic string models containing GUT gauge symmetries as subgroups, e.g.\
$E_8 \supset E_6 \supset SO(10) \supset SU(5) \supset G_\text{SM}$.

In Section~\ref{sec:theory} we will describe the theoretical framework of Schrek-Schwarz compactification, including the breaking of supersymmetry and the GUT symmetry, and discuss the placement of fermionic matter. We will describe our methodology for investigating these models in Section~\ref{sec:constraints} as well as list our low energy experimental constraints. Following this we will go on to discuss the results for each model in Sections \ref{sec:bhnmodel} to \ref{sec:e6}. These will include the Barbieri, Hall and Nomura $SU(5)$ model~\cite{Barbieri1} in Section~\ref{sec:bhnmodel}; an $SU(5)$ model with an additional singlet in Section~\ref{sec:su5singlet}; an $SU(5) \times U(1)$ model in Section~\ref{sec:su5xu1}; and an $E_6$ model in Section~\ref{sec:e6}. In Section~\ref{sec:conclusions} we will summarise our findings and draw some conclusions.

\section{Theoretical Framework} \label{sec:theory}

Here, we briefly review the theoretical framework of Schrek-Schwarz models. In Section~\ref{sec:SScompactification} we will discuss the compactification and introduce our additional symmetries, and show how this may be projected onto 4 dimensions in Section~\ref{sec:projection}. We will demonstrate how this can be used to break supersymmetry and the unification gauge group in Sections~\ref{sec:susybreak} and \ref{sec:gaugebreak} respectively. Finally we will discuss the placement of fermionic matter in Section~\ref{sec:fermionic}.

\subsection{Scherk-Schwarz compactification}
\label{sec:SScompactification}

We first briefly review the Schrek-Schwarz compactification of extra dimensions, following the notation of Quiros \cite{Quiros}, to provide context for our study and set our notation. Here we initially consider only 5-dimensional models and restrict ourselves to flat compactifications. We split our space-time coordinates into $x^\mu$, defined on our usual flat Minkowski space-time, and $y$, our extra coordinate describing the compactified space $C$ of finite size $R$. We work in the regime $E \ll 1/R$ and integrate out the heavy modes of the theory, resulting in a 4-dimensional effective field theory of the 5-dimensional action.

In general, the compact manifold $C$ may be written in terms of a non-compact manifold $\mathcal{M}$, modded out by a discrete group $G$, so that $C = \mathcal{M} / G$. The discrete group $G$ acts freely on the manifold $\mathcal{M}$ via some operators $\tau_g$,
\begin{equation}
 \tau_g : \mathcal{M} \rightarrow \mathcal{M},	\qquad g \in G,
\end{equation}
where the $\tau_g$ live in the representation space of $G$. The compact space is obtained by identifying points that belong to the same `orbit',
\begin{equation}
 y \equiv \tau_g (y),
\end{equation}
which in turn must be reflected in the symmetry of our theory. That is, physics should not be dependent on individual points in $y$, but rather on their orbits, and our (5-dimensional) Lagrangian must reflect this identification,
\begin{equation}
 \mathcal{L}_5 [\phi(x^{\mu}, y)] = \mathcal{L}_5 [\phi(x^{\mu}, \tau_g(y))],
\end{equation}
where $\phi(x^\mu,y)$ are some generic fields.

Clearly a {\em sufficient} condition on these fields is $\phi(x^{\mu}, \tau_g(y)) = \phi( x^{\mu},y)$, which leads to what we call {\em ordinary compactification}. However, a more general {\em necessary and sufficient} condition is $\phi(x^{\mu}, \tau_g(y)) = T_g \phi( x^{\mu}, y )$, where $T_g$ is an appropriate representation of $G$ acting on field space. The case $T_g = 1$ recovers ordinary compactification, but non-trivial $T_g$ results in {\em Scherk-Schwarz compactification}, which is the main focus of this paper.

The simplest compact space we can use is the circle, $C = S^1$, which may be constructed as the identification $\mathds{R}^1/\mathds{Z}$, where $\mathds{R}^1$ is the real line and $\mathds{Z}$ corresponds to a translation of $2 \pi n R$ with $n \in \mathds{Z}$. The action of the infinite discrete group modulo $\mathds{Z}$ is given by the operators $\tau_n$, acting on elements $y \in \mathds{R}^1$ by mapping them onto
\begin{equation}
 \tau_n (y)  = y + 2\pi n R, \qquad n \in \mathds{Z},
\end{equation}
where $R$ is the radius of the circle $S^1$. Effectively we've taken the real number line $\mathds{R}^1$ and `curled' it up, therefore restricting the domain of our manifold to $[y, y+ 2\pi R)$. The first generator of $\mathds{Z}$,  $\tau_1 = 2\pi R$, will correspond to the only independent `twist' $T$, acting on fields $\phi$ according to,
\begin{equation}
 \phi (x^{\mu}, \tau_1(y)) = \phi (x^{\mu}, y + 2\pi R) = T \phi(x^{\mu}, y), \label{PeriodicT}
\end{equation}
since the other operators $\tau_n$, $n>1$, can be built out of multiple applications of $\tau_1$.

\begin{figure}[h]
 \begin{center}
  \includegraphics[scale = 0.4]{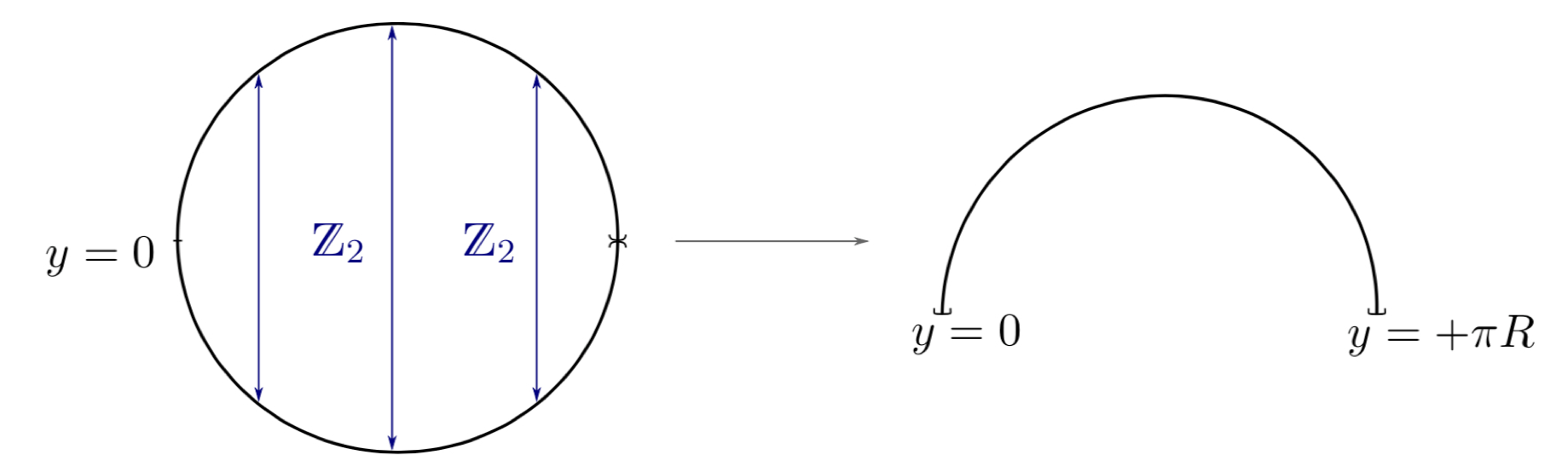}
  \caption{Modding $S^1$ by $\mathds{Z}_2$}\label{fig:CircleFold}
 \end{center}
\end{figure}

Unfortunately these 5-dimensional models do not allow for chiral fermions~\cite{Sundrum}, due to the 5D Lorentz algebra containing $\gamma^5$ resulting in the smallest irreducible representation being a Dirac fermion. To overcome this we further `fold' our extra dimension, converting our circle into an interval. This is an {\em orbifold} compactification. We assign a parity to our fields under a $\mathds{Z}_2$ transformation
\begin{equation}y \to \xi(y)=-y, \end{equation}
which identifies the lower half of our circle with the upper half, as seen in Figure~\ref{fig:CircleFold}. The manifold $O = S^1/\mathds{Z}_2$ is no longer smooth but now becomes an {\em orbifold} with {\em fixed points} at $y=0$ and $\pi R$.

As before, our Lagrangain must remain invariant under this transformation, so the fields transform as
\begin{equation}
 \phi( x^{\mu}, \xi_h (y) ) = Z \phi (x^{\mu}, y),
\end{equation}
where $Z$ is the parity assignment. When integrating over the 5th dimension and writing our theory as a 4-dimensional effective theory, we will generate an effective tower of  Kaluza-Klein states. Under the $\mathds{Z}_2$ identification the two component spinors within the Dirac fermion will have opposite parities \cite{Csaki:2005vy}. Those which have an even assignment (i.e.\ von Neumann boundary conditions) will be allowed to have zero modes whereas the odd ones (i.e.\ Dirichlet boundary conditions) will not.

Therefore, by choosing our parities we can prevent whichever zero mode we like from appearing, allowing us to lift the right-handed fermions and regain, what is in effect, a chiral theory.

We note that $Z^2 = \mathbbm{1}$, and $Z, T$ must obey the \emph{ consistency condition},
\begin{equation}
 TZT=Z \quad \Leftrightarrow \quad ZTZ=T^{-1} \label{consistencyCond}
\end{equation}
We can easily see this latter relation geometrically by applying  $y \xrightarrow{\tau} y+ 2 \pi R \xrightarrow{\xi} - y - 2 \pi R \xrightarrow{\tau} -y = \xi(y),$ and requiring an analogous relation between the field operators.

Recall that $T$ corresponds to an operator expressing the symmetry defined by $G$, so we can write $T$ as,
\begin{equation}
 T = \exp( 2\pi i \beta^a \lambda^a),
\end{equation}
where $\beta^a$ parameterises the symmetry transformation, and $\lambda^a$ are the Hermitian generators. For infinitesimal transformations we may rewrite the consistency condition to $\mathcal{O}(\beta^2)$ as,
\begin{equation}
 \{ \beta^a\lambda^a, Z \} = 0	\label{ZTantiCommute}.
\end{equation}

We will later be interested in fields that transform as doublets under a global SU(2) symmetry. In this case since $Z^2 = \mathbbm{1}$, we have two choices for $Z$, i.e.\ $Z = \sigma_3$ or $Z = \pm \mathbbm{1}$. The latter choice $Z=\pm \mathbbm{1}$ requires $T = \pm \mathbbm{1}$ and we recover ordinary compactifications. For the non-trivial case, $Z = \sigma_3$ and the generators of $T$ are also Pauli matrices, $\vec{\lambda} = \vec{\sigma}$. If $T = \exp(i\pi\sigma^3) = \mathbbm{1}$, we again have an ordinary compactification. The remaining solution $T = \exp( 2\pi i (\beta^1 \sigma^1 + \beta^2 \sigma^2))$ may be simplified by rotating away the $\sigma^1$ direction whilst preserving $\sigma^3$, so that,
\begin{equation}
 Z = \sigma^3,			\qquad			T = \exp ( 2\pi i \alpha \sigma^2 ),
\end{equation}
where $\alpha$ parameterises the transformation.

\subsection{Projecting to 4 dimensions}
\label{sec:projection}

The building blocks of our 5D theory are {\em hypermultiplets} $\mathscr{H}$, and {\em vector multiplets} $\mathscr{V}$, which we present in the formalism introduced by Mirabelli and Peskin \cite{Mirabelli}. The hypermultiplets $\mathscr{H}$ consist of complex scalars $A^i$, $i =1, 2$ and a Dirac spinor $\Psi$,
\begin{equation}
        \mathscr{H} = (A^i, \Psi).
      \end{equation}
Note that the minimal fermionic matter field in 5D is the Dirac spinor since it is the lowest weight representation of the Lorentz algebra. Furthermore note that $\Psi, A^i$  transform as a doublet under $SU(2)_R$ \cite{Sohnius}, the residual 5D supersymmetry. The Vector multiplets $\mathscr{V}$ consist of the 5D gauge fields $A_M, \mbox{ M = 0\ldots 5}$, gauginos $\lambda^i$, $i= 1 ,2$, and a scalar $\Sigma$ in the adjoint representation,
\begin{equation}
        \mathscr{V} = (A_M, \lambda^i, \Sigma).
      \end{equation}
$\lambda^i$ transforms as a doublet under $SU(2)_R$, where $\lambda^i$ are symplectic Majorana spinors,
\begin{equation}
        \lambda^i = \begin{pmatrix}	 \lambda^i_L \\ \vspace*{-0.25cm}\\ \epsilon_{ij}\overline{\lambda}_{jL} \end{pmatrix},
        \qquad	\overline{\lambda}_{jL} \equiv -i \sigma^2 (\lambda^j_L)^*,
      \end{equation}
with $\lambda^i_L$ a left handed Weyl spinor.
These are defined in the 5D space-time with,
\begin{equation}
 \eta_{MN} = \diag(1, -1, -1, -1, -1), \qquad \gamma^M = \{ \gamma^{\mu}, \gamma^5 \}, \qquad
 \gamma^5 = \begin{pmatrix}	-i & 0 \\
 0 & i
 \end{pmatrix} \otimes I_2,
\end{equation}
and
\begin{equation}
 \sigma^{\mu} = (\mathbbm{1} , \vec{\sigma}), \qquad \overline{\sigma}^{\mu} = ( \mathbbm{1}, -\vec{\sigma}),
\end{equation}
where again we emphasize that we're considering a flat extra dimension, ignoring the brane tension (i.e.\ {\em not} a warped scenario).

The on-shell vector multiplet $\mathscr{V}$ is extended to off-shell by adding a $SU(2)_R$ triplet of real valued auxiliary fields $X^a, a= 1,2,3$, and the hypermultiplet $\mathscr{H}$ is similarly extended by adding a complex doublet of auxiliary fields $F^i, i=1, 2$,
\begin{align}
 \mathscr{V}_{\text{on-shell} } &= (A_M, \Sigma, \lambda^i) 	&&\rightarrow	&	\mathscr{V}_{\text{off-shell }} &= (A_M, \Sigma, \lambda^i, X^a),	\\
 \mathscr{H}_{\text{on-shell}} &= (A^i, \Psi)		&&\rightarrow	&		\mathscr{H}_{\text{off-shell}} &= (A^i, \Psi, F^i).
\end{align}
These fields obey the supersymmetry transformations of Ref.~\cite{Quiros}.

For the $S^1 / \mathds{Z}_2$ orbifold, the fixed points at $y =0$ and $\pi R$ provide 4-dimensional Minkowski manifolds, and compactification will result in a tower of Kaluza-Klein states as usual. We may restrict which zero-modes appear by assigning their $\mathds{Z}_2$ assignment to $+1$ if the zero-mode is to be allowed and $-1$ if we want to forbid it. Specifically, by choosing $Z = \sigma^3$, we assign,
\begin{align}
 Z = +1 & : 			\qquad	A^M, \lambda^1_L, X^3 ; \qquad \quad \hspace{0.15cm} \xi^1_L, \\
 Z = -1 & :			\qquad	A^5, \Sigma, \lambda^2_L, X^{1,2} ; \qquad \xi^2_L,
\end{align}
where $\xi^i, i=1,2$ are the corresponding 5D supersymmetry parameters, which are symplectic Majorana spinors.

After orbifolding, the states on the $y = 0$ brane will obey reduced supersymmetric transformations \cite{Quiros}, and
\begin{equation}
 \begin{rcases}
  \delta_\xi X^3 = (\xi^1_L)^\dagger \overline{\sigma}^{\mu} \mathcal{D}_\mu \lambda^1_L  - i (\xi^1_L)^\dagger \mathcal{D}_5 \overline{\lambda}^2_L + h.c. \\
  \delta_\xi (\partial_5 \Sigma) = -i (\xi^2_L)^\dag \mathcal{D}_5 \overline{\lambda}^2_L + h.c.
\end{rcases} \quad \Rightarrow \quad \delta_\xi (X^3 - \partial_5 \Sigma) = \xi^1_L \overline{\sigma}^\mu \mathcal{D}_{\mu} \lambda^1_L  + h.c.
\end{equation}
We see that $X^3 - \partial_5 \Sigma$ transforms as a total derivative. The vector multiplet projected onto the brane is then, in the Wess-Zumino gauge, $(A^{\mu} , \lambda^1_L, D)$	where $D =X^3 - \partial_5 \Sigma$ \cite{Hebecker:2001ke, ArkaniHamed:2001tb}.

Analogously for the hypermultiplet, starting with $\mathcal{H} = (A^i, \Psi, F^i)$, with Dirac spinor  $\Psi = ( \psi_L, \psi_R ) $, we have the assignment,
\begin{align}
 Z = +1 & : 			\qquad	A^1, \psi_L, F^1; \qquad \xi^1_L, \\
 Z = -1 & :			\qquad	A^2, \psi_R, F^2; \qquad \xi^2_L.
\end{align}
After orbifolding we have,
\begin{equation}
 \begin{rcases}
  \delta_\xi F^1 = i\sqrt{2} (\xi^1_L)^\dag \overline{\sigma}^{\mu} \partial_{\mu} \psi_L + \sqrt{2}(\xi^1_L)^\dag 						\partial_5\psi_R \\
  \delta_\xi (\partial_5 A^2) = \sqrt{2} (\xi^1_L)^\dag \partial_5 \psi_R
 \end{rcases} \quad \Rightarrow \quad \partial_\xi(F^1 - \partial_5 A^2) = i\sqrt{2} (\xi^1_L)^\dag \overline{\sigma}^{\mu} 					  \partial_{\mu} \psi_L
\end{equation}
which also transforms as a total derivative. The off-shell chiral supermultiplet on the $y= 0$ brane is $(A^1, \psi_L, F)$ with $F = F^1 - \partial_5 A^2$.

The 5D Lagrangian for the gauge fields is the standard 5D Super Yang-Mills Lagrangian \cite{Hebecker:2001ke},
\begin{equation}
 \mathcal{L}_5 = \Tr \left\{ -\frac{1}{2} F^2_{MN} + (\mathcal{D}_M \Sigma)^2 + i\overline{\lambda} \gamma^M \mathcal{D}_M 							\lambda + \vec{X}^2 - \overline{\lambda} [\Sigma, \lambda] 	\right\}.
\end{equation}
The corresponding Lagrangian on the $y=0$ brane will have a standard form corresponding to a 4D chiral multiplet coupled to a gauge multiplet (see Quiros \cite{Quiros}). This gives a bulk and a brane Lagrangian with the added feature of a superpotential $W$ that connects the bulk and brane matter fields via the interaction of chiral superfields on the $y = 0$ brane, $W(\Phi_0, \mathcal{A})$, where by $\Phi_0$ we mean any general 4D chiral superfield.

The 5D Lagrangian for the hypermultiplet $\mathscr{H}$ components, ignoring the gauge coupling for now, will be,
\begin{equation}
 \mathcal{L}_5 = |\partial_M A^i|^2 + i\overline{\psi} \gamma^M \partial_M \psi + |F^i|^2.
\end{equation}
The brane Lagrangian involving interactions with matter will then be given by,
\begin{equation}
 \mathcal{L}_4 = F^1 \frac{\partial W}{\partial A^1} + h.c. = (F^1 - \partial_5 A^2) \frac{\partial W}{\partial A^2} + h.c.
\end{equation}
Integrating out the auxiliary field $F^1$ leaves the action,
\begin{equation}
 S = \int d^4 x \hspace{0.05cm} dy \left\{  |\partial_M A^i|^2 + i\overline{\psi} \gamma^M \partial_M \psi -
 \delta(y) \left[ (\partial_5 A^2 \frac{\partial W}{\partial A^1} + h.c.) + \delta(y) \abs{\frac{\partial W}{\partial A^1}} \right] \right\}.	\label{SuperPotential}
\end{equation}

\subsection{Supersymmetry breaking} \label{sec:susybreak}
We  first demonstrate supersymmetry breaking in the simpler case where gauge symmetry is unaffected. Consider a vector multiplet $\mathscr{V} = (A_M, \lambda^i, \Sigma)$ and two Higgs matter hypermultiplets $\mathscr{H^a} = (H^a_i, \Psi^a) \hspace{0.1cm}, a=1, 2$ which can be rotated into one another under an $SU(2)_H$ flavor symmetry. The 5D action will then be invariant under $SU(2)_R \times SU(2)_H$ with a Lagrangian,
 \begin{align}
  \mathcal{L}_5 = \displaystyle \frac{1}{g^2} \Tr
  \left\{ \lefteqn{\phantom{\frac{1}{2}}} \right. &
  -\frac{1}{2} F^2_{MN} + (\mathcal{D}_M \Sigma)^2 + i \overline{\lambda}_i \gamma^M \mathcal{D}_M \lambda^i - \overline{\lambda}_i [\Sigma, \lambda^i] + \abs{\mathcal{D}_M H^a_i}^2 +  \overline{\Psi}_a (i\gamma^M \mathcal{D}_M - \Sigma)\Psi^a  \nonumber \\
  & \left. \displaystyle
  - (i \sqrt{2}(H^a_i)^\dag \overline{\lambda}_i \Psi^a + h.c.)
  - (H^a_i)^\dag \Sigma^2 H^a_i
  - \frac{g}{2} \left( (H^a_i)^\dag \vec{\sigma}^j_i T^A H^a_j  \right)^2
  \right\}, \label{BIGLAG}
\end{align}
as long as the fields are of the appropriate respresentations, e.g.\ $\lambda^i \sim (\mathbf{2}_R, 1_H)$, $\Psi^a \sim (1_R, \overline{\mathbf{2}}_H)$, $H^a_i \sim (\mathbf{2}_R, \overline{\mathbf{2}}_H)$, with the subscripts $R, H$ refering to $SU(2)_R$ or $SU(2)_H$.

With our choice of $Z = \sigma^3$ we then have the eigenvalues,
\begin{align}
 Z = +1 & : 			\qquad	\lambda^1_L, V_{\mu};	\qquad H^1_1, \psi^1_R; \quad H^2_2, \psi^2_L,                  \\
 Z = -1 & :			\qquad	\lambda^2_L, V_5, \Sigma;	\hspace{0.09cm}\quad H^1_2, \psi^1_L;	\quad H^2_1, \psi^2_R,
\end{align}
which forbids massless Kaluza-Klein modes  for the $Z=-1$ states. The parity operator may be written as a product of operators acting on either the $SU(2)_R$ or $SU(2)_H$ symmetries,
\begin{equation}
 Z = \pm (\sigma^3)_R \otimes (\sigma^3)_H \otimes i\gamma^5,
\end{equation}
where the $i\gamma^5$ acts only on the spinor indices of the representations to project the left/right handed chirality of the Dirac spinors.

Extending the twist operator $T$ to $SU(2)_R \times SU(2)_H$ gives,
\begin{equation}
 T = e^{2\pi i \alpha \sigma^2} \otimes - e^{ 2\pi i \gamma \sigma^2},
\end{equation}
where $\alpha$ parameterises the $SU(2)_R$ symmetry, and $\gamma$ the $SU(2)_H$. Under this twist, fields $\phi$ must obey our boundary conditions,
\begin{equation}
 \phi (x^{\mu}, y+2\pi R) = e^{ 2\pi i \alpha \sigma^2 } \phi(x^{\mu}, y),
\end{equation}
where to illustrate the argument we've just taken the action dictated by the $SU(2)_R$ field space. The above has the trivial solution:

\begin{equation}
 \phi(x^{\mu}, y+ 2\pi R) = e^{i \alpha \sigma^2 y /R} \tilde{\phi} (x^{\mu}, y)
\end{equation}
where $ \tilde{\phi} (x^{\mu}, y + 2\pi R) = \tilde{\phi} (x^{\mu}, y)$ is a periodic field in $y$ and can be in turn be expanded into its KK modes.

Applying this reasoning to our fields, we find,
\begin{align}
 \begin{pmatrix}	\lambda_1 \\ \lambda_2	\end{pmatrix} &= e^{i \alpha \sigma^2 y /R}  \begin{pmatrix}	\tilde{\lambda}_1 \\ \tilde{\lambda}_2	\end{pmatrix},			\\[2mm]	\begin{pmatrix}\Psi^1 \\ \Psi^2	\end{pmatrix} &=  \begin{pmatrix}	\tilde{\Psi}^1 \\[2mm] \tilde{\Psi}^2	\end{pmatrix}  e^{-i \gamma \sigma^2 y /R}, \\[2mm]
 \begin{pmatrix} H^1_1 & H^1_2 \\[1mm] H^2_1 & H^2_2	\end{pmatrix} &= e^{i \alpha \sigma^2 y /R} \begin{pmatrix} \tilde{H}^1_1 & \tilde{H}^1_2 \\ \tilde{H}^2_1 & \tilde{H}^2_2	\end{pmatrix}  e^{-i \gamma \sigma^2 y /R},
\end{align}
where each aquires an $\alpha$ and/or $\gamma$ parameterised exponential according to their transformation properties under $SU(2)_R \times SU(2)_H$.

Applying this to the Lagrangian of Eq.~(\ref{BIGLAG}), the kinetic part, or more specifically the $\partial_5$ derivative, acts on the boundary conditions giving us effective $4D$ soft SUSY breaking masses as in Barbieri, Hall, and Nomura's model \cite{Barbieri1},
\begin{equation}
 \mathcal{L}_{\cancel{SUSY}} = - \frac{1}{2} \frac{\alpha}{R} (\lambda^{1 (0)}_L \lambda^{1 (0)}_L + h.c.)
                               -\left( \frac{\alpha^2}{R^2} + \frac{\gamma^2}{R^2} \right) (\abs{h_u}^2 + \abs{h_d}^2)  +  \frac{2\alpha\gamma}{R^2}(h_u h_d + h.c)
                               - \frac{\gamma}{R}(\overline{\psi}_h \psi_h + h.c.),
\end{equation}
where we've labeled the zero-modes of our solutions, $h_u = H^{1(0)}_1$, $h_d = H^{2(0)}_2$, $\overline{\psi}_h = \overline{\psi}^{2(0)}_L$, $\psi_h = \psi^{1(0)}_R$. In the language of the MSSM, the Scherk-Schwarz  twists have generated universal gaugino breaking terms ($m_0 = \hat{\alpha}$), and holomorphic Higgs terms ($ m^2_{H_u} = m^2_{H_d} = \hat{\alpha}^2$,  $\mu  = \hat{\gamma}$, $\mu B = -2 \hat{\alpha}\hat{\gamma}$) via the $\hat{\alpha} \equiv \alpha/R, \hat{\gamma} = \gamma / R$ parameters controlling the $SU(2)_R \times SU(2)_H$ breaking.

\subsection{Gauge Breaking} \label{sec:gaugebreak}
We have seen how the Scherk-Schwarz compactification provides supersymmetry breaking, but it can also break our GUT's gauge symmetry $\mathscr{G}$ to a subgroup $\mathscr{H}$ on the brane. To do this, we extend the definition of the parity assignment on the fields with non-trivial gauge structure to,
\begin{align}
 A_M^A (x^{\mu}, -y) &= \alpha^M \Lambda^{AB} A^B_M (x^{\mu}, y ), \\
 \psi (x^{\mu}, -y) &= \lambda_R \otimes (i\gamma^5) \psi (x^{\mu},  + y),
\end{align}
where $\alpha^M = \pm 1$ are the previous parity assignments, $\Lambda^{AB}$ is a matrix with $\Lambda^2 = 1$ and eigenvalues $\pm 1$, and $\lambda_R$ is a hermitian matrix acting on the representation space of the field $\psi_R$. In order to keep the bulk kinetic term $F^A_{MN} F^{A \, MN}$ invariant, $\Lambda$ must satisfy,
\begin{equation}
 f^{ABC} = \Lambda^{AA'} \Lambda^{BB'} \Lambda^{CC'} f^{A'B'C'},
\end{equation}
where $f^{ABC}$ are the structure constants of the gauge group. Since $\Lambda$ has eigenvalues $\pm 1$ it can be written in a diagonal basis as $\Lambda^{AA'} = \delta^{AA'} \eta^{A'}$, with $\eta^{A'} = \pm 1$. In this basis we have,
\begin{equation}
 f^{ABC} = \eta^A \eta^B \eta^C f^{ABC},
\end{equation}
where there is no summation over repeated indices. We are free to choose whatever parity assignment $\eta$'s we like, and break the gauge symmetry, as long as they obey this constraint. Conversely, setting all $\eta$'s to $1$ recovers the trivial case of $\Lambda=1, \lambda_R = 1$, maintaining the gauge symmetry. To break our group $\mathscr{G}$ to a subgroup $\mathscr{H}$ we must therefore keep the parities of field components in the directions corresponding to the generators of $\mathscr{H}$ even, while setting the others to be odd.

We simplfy the treatment by choosing the $\eta^A$'s such that the generators $T^A$ are naturally split into two cases. Firstly, $T^a$ with $\eta^a = +1$ such that the surviving gauge group has generators $\mathscr{H} = \{ T^a \}$. These $T^a$ transforms as $T^a \rightarrow \delta^{aa'} \eta^{a'} T^{a'} = T^{a}$ so that the automorphism and the subgroup is preserved. Secondly, $T^{\hat{a}}$ with $\eta^{\hat{a}} = -1$ such that the broken group has generators
$\mathscr{K} = \mathscr{G} / \mathscr{H} = \{ T^{\hat{a}} \}$ (and now $T^{\hat{a}}\rightarrow - T^{\hat{a}}$). For example if our gauge group is $SU(2)$ and we choose $a= 3; \hat{a} = 1,2$ we would have $SU(2)$ breaking down to $U(1)$.

The $\eta$ assignment will also impact the fields that live in the gauge representation space. Since we require the bulk action be invariant, we require the coupling,
\begin{equation}
 i g A^{A}_M \overline{\psi} \gamma^M T^A \psi
\end{equation}
remain invariant. To achieve this the $\lambda_R$ matrix must satisfy,
\begin{equation}
 [ \lambda_R, T^{a}_R ] = 0 \qquad \{ \lambda_R, T^{\hat{a}}_R \} =0
\end{equation}
Our choice in $\Lambda$ has split our representation into two implicit subspaces, with the $Z$ parity assignment dictated by the (anti-)commutation relations.

For example, taking $SU(5)$ as the unification gauge group, we may choose $\Lambda$ such that $T^a \in G_{SM}, T^{\hat{a}} \in SU(5)/ G_{SM}$ so that $\lambda_R = \diag(+1, +1, +1, -1, -1)$, and the lowest non-trivial $SU(5)$ representation, $\mathbf{5}$, is naturally separated into $\mathbf{3} \oplus \mathbf{2}$. Fields with $Z = -1$ are prevented from having zero-modes, and acquire a heavy mass of $\mathcal{O}(1/R)$ via the $\partial_5$ derivative. The surviving gauge group can use the standard Higgs mechanism to undergo the usual Standard Model electro-weak breaking.

We noted earlier that we can combine our $Z$ and $T$ transformations to form an alternative $Z'$, giving us the equivalent orbifold $\mathds{R}^1 / \mathds{Z}_2 \times \mathds{Z}_2'$. The above gauge breaking argument may be applied to the $\mathds{R}^1 / \mathds{Z}_2 \times \mathds{Z}_2'$ orbifold. To this extent the gauge breaking can be assigned to either  $Z, Z'$ or the translation $T$ or a combination (due to them being isometries obeying the consistency condition in Eq.~\ref{consistencyCond}).

The physical symmetry of the theory then consists of the generators $T^a$ that simultaneously commute with the chosen forms for  $Z, Z',T$.
 If we take $Z \sim \diag(+, +, +, +, +)$ and want to achieve  $SU(5) \rightarrow G_{SM}$ breaking, we can choose,
\begin{equation}
 Z \sim \diag(+, +, +, +, +), \qquad T \sim \diag (+,+,+,-,-), \label{ZTmod}
\end{equation}
which was explored originally by Kawamura \cite{Kawamura:2000ev, Kawamura:1999nj}.

Note that the simultaneously anti-commuting generators $T^{\hat{a}}$ will determine the presence of non-trivial Wilson lines  phases which can lead to Hosotani breaking \cite{Hosotani1, Hosotani2, Hosotani3}, depending on the matter content of the theory. The above form of the gauge symmetry breaking assignment is chosen to ensure that we do not have any Wilson line phases present in the 4D theory.

To summarise, the actions of our isometries on field space are defined by,
\begin{align}
 Z &= (\sigma^3)_R \otimes (\sigma^3)_H \otimes \diag(+,+,+,+,+), \\
 T &= e^{2\pi i \alpha \sigma^2} \otimes -e^{ 2\pi i \gamma \sigma^2} \otimes \diag (+,+,+,-,-).
\end{align}
The Scherk-Schwarz compactification allows us to break both supersymmetry and the unification gauge group on the $y=0$ brane.

\subsection{Fermionic Matter: Brane vs Bulk} \label{sec:fermionic}

The fermionic matter allows some freedom in whether they should be in the bulk as hypermultiplets via $\mathcal{L}_5$ or only on the brane as chiral multiplets via $\mathcal{L}_{4i}, i=0$. Their placement will impact the number of required multiplets to provide the low energy Standard Model fields. For clarity, in this discussion we will assume an $SU(5)$ gauge structure.

We begin with the simplest placement, brane matter. In this case we use the usual chiral multiplets from an ordinary $SU(5)$ model, i.e.\ the supersymetric Standard Model fields $U, D, Q, L, E$ which are contained in the $T_{\mathbf{10}} \sim \mathbf{10} \supset \{ Q, U, E\} $ and $ F_{\mathbf{\overline{5}}} \sim \overline{\mathbf{5}} \supset \{ D, L \}$. These representations are now coupled to the $\mathds{Z}_2$ chiral projection of the Higgs hypermultiplets on the brane via the superpotential $W$.

We note that when projecting the bulk matter hypermultiplets we form two chiral multiplets defined by either $Z = \pm 1$. The components of the hypermultiplet must transform to maintain gauge invariance in the bulk as dictated by the Lagrangian in Eq.~(\ref{BIGLAG}). More specifically the components contained in the $Z = +1$ chiral multiplet will transform as the fundamental of the group while those in the $Z = -1$ one will transform as the conjugate, which we denote with a superscript $^c$. For an arbitrary matter hypermultiplet, coupled to an $SU(5)$ gauge structure,
\begin{equation}
 \mathscr{A} = (A^i, \Psi_a) \qquad \rightarrow \qquad \mathcal{A} = (A^1, \psi^A_R) \sim \mathbf{5}; \quad \mathcal{A}^c = (A^2, \psi^A_L) \sim \overline{\mathbf{5}}
\end{equation}
Therefore as usual we have,
\begin{equation}
 S_{\text{Matter}} = \int d^4 x \hspace{0.1cm} dy \hspace{0.1cm} \delta(y) \left[ \int d^2 \theta \sum_{j, k =1}^3 (y_1)_{jk} T_{\mathbf{10}_j} T_{\mathbf{10}_k} H^c_{\mathbf{5}} + (y_2)_{jk} T_{\mathbf{10}_j} F_{\mathbf{\overline{5}}_k} H_{\mathbf{\overline{5}}} + h.c.\right]
\end{equation}
where $H_{\mathbf{5}} = (H^1_1, \psi^1_R), H_{\mathbf{\overline{5}}} = (H^2_2, \psi^2_L)$, and we've introduced $3$ generations denoted by the index structure $j, k$. After orbifolding, the $H_{\mathbf{5}}, H_{\mathbf{\overline{5}}}$ automatically acquire a $2-3$ splitting and the rest of the model's phenomenology is analogous to the usual supersymetric SU(5) GUT.

If on the other hand we put our matter fields as components of hypermultiplets in the bulk we run into another issue. Since all the bulk hypermultiplets will automatically undergo the $2-3$ splitting induced by the $T$ action, inserting just one of the chiral analogs $\mathscr{T}_{\mathbf{10}}$, $\mathscr{F}_{\mathbf{\overline{5}}}$ would result in having some of the states in the Standard Model spectrum projected out, i.e.\ we would not have the correct zero-mode spectrum. To get around this we must add two copies of each $SU(5)$ fermionic matter hypermultiplet, assigned opposite $Z$ parities with respect to each other. That is, we introduce $4$ hypermultiplets
$\mathscr{T}_{\mathbf{10}} = \{ T_{\mathbf{10}}, T^c_{\mathbf{10}} \}$,
$\mathscr{T}'_{\mathbf{10}} = \{ T'_{\mathbf{10}}, T'^c_{\mathbf{10}} \}$,  $\mathscr{F}_{\mathbf{\overline{5}}}  = \{  F_{\mathbf{\overline{5}}},  F^c_{\mathbf{\overline{5}}} \}$,
$\mathscr{F}'_{\mathbf{\overline{5}}} = \{  F_{\mathbf{\overline{5}}},  F^c_{\mathbf{\overline{5}}} \}$, which we give the $Z$ assignments,
\begin{align}
 \{ T_{\mathbf{10}}, T^c_{\mathbf{10}} \} &\rightarrow \{ (+) T_{\mathbf{10}}, (-) T^c_{\mathbf{10}} \}, \\
 \{ T'_{\mathbf{10}}, T'^c_{\mathbf{10}} \} &\rightarrow \{ (-) T'_{\mathbf{10}}, (+) T'^c_{\mathbf{10}}, \}
\end{align}
and analogous assignments for $\mathscr{F}_{\mathbf{\overline{5}}}$, $\mathscr{F}'_{\mathbf{\overline{5}}}$. With these assignments our Lagrangian becomes that presented in Ref.~\cite{Barbieri1}.

However with this matter placement we have another added complexity since the individual hypermultiplets transform under the residual $SU(2)_R$ symmetry (note that we assume a trivial flavour action acting on $\mathscr{T}, \mathscr{F}$ ). After orbifolding, the non-trivial SS conditions provide us with squark soft SUSY breaking masses via the kinetic part of the Lagrangian in Eq.~(\ref{BIGLAG}), along with a contribution to the trilinear squark coupling $A_0$ via the $\partial_5 Q^2$ term in Eq.~(\ref{SuperPotential}).

\section{Methodology and Constraints}
\label{sec:constraints}

The compactification of the high scale extra dimensional model provides us with an effective 4D softly broken supersymmetric model at high energies. We would like to examine this model's low energy spectrum to ensure that it is phenomenologically consistent with experimental observations. We include as inputs the high scale model parameters and use these to set the soft SUSY breaking parameters. We then use RGEs to evolve our parameters down to the low scale, where we apply constraints. For a study of Scherk-Schwarz with an electroweak scale compactification, see Ref.~\cite{Garcia:2015sfa}.

The RGE running is performed using the FlexibleSUSY [v.2.0.1] \cite{FlexibleSUSY, Athron:2017fvs} spectrum generator which uses numerical routines generated by SOFTSUSY \cite{Allanach:2001kg,Allanach:2013kza} and with two-loop RGEs provided by SARAH [v.4.12.2] \cite{Staub:2009bi,Staub:2010jh,Staub:2012pb,Staub:2013tta}. SARAH also provides the electroweak tadpole conditions. For example, in the $SU(5)$ model discussed in Section~\ref{sec:bhnmodel} the high scale inputs are $\hat{\alpha}$ and $\hat \gamma$, which we relate to the soft SUSY breaking parameters via Eqs.~(\ref{eq:soft_masses_SU5}) and (\ref{eq:squark_masses_brane}), and these are then run down to the low scale where electroweak symmetry is broken and experimental constraints applied.

In principle, the electroweak tadpole equations could set our final low energy observables, the ratio of vacuum expectation values (vevs) of the two Higgs doublets, $\tan\beta$, and the $Z$-boson mass. However, for technical reasons it is easier to assign these values at the low scale. This means we have to (temporarily) relax some of our high scale relations between the soft SUSY breaking parameters and the model inputs. We choose to allow our choice of $\tan \beta$ to fix $\hat \gamma$ and leave $\mu B$ (the soft SUSY breaking parameter corresponding to the Higgs-higgsino mass parameter) unfixed. Only at the end of the process will we check if $\mu B = -2 \hat \alpha \hat \gamma$ as required by Scherk-Schwarz compactification. We will refer to this as the `Scherk-Schwarz condition'. In practice, we will not insist this condition is obeyed exactly, due to the uncertainties arising from the RGE running. Instead we will insist that the Scherk-Schwarz condition is obeyed with 95\% confidence. We stress that in principle, this is no different than forcing the relation at high energies and searching for values of $\tan \beta$ that satisfy the tadpole equations.

To explore the parameter space we employ a `seeded random walk' scanning algorithm. We first sample the phase space with a uniform distribution to find points that produce EWSB and inspect if they come close to satisfying our required constraints (such as the correct Higgs mass), with `closeness' being defined by a global $\chi^2$. Then we perform a random walk around each point to search for those with a better fit and if such a point is found it becomes the new seed. This is repeated until we find a point that agrees with the required constraints (if it exists). The search is abandoned if computation time exceeds a preset limit.  This provides us with points that are theoretically well behaved but may still be experimentally excluded. We therefore then must check LHC and Dark Matter constraints.

We apply LHC bounds and constraints from the ATLAS and CMS collaborations:
\begin{enumerate}
 \item We insist on a Higgs mass in the range $123 \leq m_H \leq 127\,$GeV, where we've assumed a $2\,$GeV theoretical uncertainty dominates those from the experimental measurement \cite{HiggsATLAS,HiggsCMS}.
 \item We require a gluino mass $m_{ \tilde{g} } \geq 2\,$TeV \cite{Aaboud:2017vwy, Sirunyan:2017pjw}.
 \item We require a lightest neutralino  and chargino masses to be outside the exclusion contour provided by Figure 13 of Ref.~\cite{ATL-PHYS-PUB-2019-022}, which in particular combines the exclusions from \cite{Aad:2015eda, Aaboud:2018jiw}. 
 \item The stop quark $m_{\tilde{t}}$ should be heavier than $1\,$TeV \cite{Aaboud:2017aeu}.
 \item Any extra gauge boson must have mass $m_{Z'} \geq 2.4\,$TeV \cite{Aaboud:2017sjh}.
\end{enumerate}

For scenarios that pass the LHC constraints and satisfy the Scherk-Schwarz constraint, we apply constraints on the Dark Matter relic density. We use the measurement from Planck \cite{PlanckData},
\begin{equation}
 \Omega_c h^2 = 0.1157 \pm 0.0023,
\end{equation}
and include a further $10\%$ uncertainty arising from the mass difference from MicrOmegas \cite{Belanger:2001fz,Belanger:2004yn,MicrOmegas} and FlexibleSUSY. We therefore accept points with a Dark Matter relic density smaller than $\Omega_c h^2 = 0.1275$ to allow for the possibility of other sources of Dark Matter.

\section{The Barbieri, Hall and Nomura SU(5) Model} \label{sec:bhnmodel}

We first model we consider is an $SU(5)$ GUT in 5D, compactified on the $S^1/Z_2$ orbifold, as proposed by Barbieri, Hall and Nomura~\cite{Barbieri1}. This model contains a vector multiplet $\mathscr{V} = (A_M, \lambda^i, \Sigma)$ and two Higgs hypermultiplets $\mathscr{H}^a = (H^a, \Psi^a)$, $a=1, 2$. The 5D action is invariant under an $SU(2)_R \times SU(2)_H$ global symmetry where the fields have the representations
$\lambda^i \sim (\mathbf{2}_R, 1_H)$,
$\Psi^a \sim (1_R, \mathbf{2}_H)$,
$H^a_i \sim (\mathbf{2}_R, \mathbf{2}_H)$.
The extra dimension is compactified at a scale $1/R = 10^{16}\,$GeV to break both the $SU(5)$ symmetry and the supersymmetry. Under the compactification symmetries $y \leftrightarrow -y$ and $y \leftrightarrow y +2 \pi R$ the fields transform with
\begin{align}
 Z &= (\sigma^3)_R \otimes (\sigma^3)_H \otimes \diag(+,+,+,+,+), \\
 T &= e^{2\pi i \alpha \sigma^2} \otimes - e^{ 2\pi i \gamma \sigma^2} \otimes \diag (+,+,+,-,-),
\end{align}
using the notation of \cite{Quiros}, where the final matrix is acting on the $SU(5)$ space.

The derivative with respect to the fifth dimension in the kinetic part of the Lagrangian acts on the boundary conditions giving us effective $4D$ soft SUSY breaking terms of the form \cite{Barbieri1}
\begin{align}
 \mathcal{L}_{\cancel{SUSY}} =&
 - \frac{1}{2} \frac{\alpha}{R} (\lambda^{1 (0)}_L \lambda^{1 (0)}_L + h.c.)
 -\left( \frac{\alpha^2}{R^2} + \frac{\gamma^2}{R^2} \right) (\abs{h_u}^2 + \abs{h_d}^2)  +  \frac{2\alpha\gamma}{R^2}(h_u h_d + h.c) \nonumber \\
&- \frac{\gamma}{R}(\overline{\psi}_h \psi_h + h.c.),
\end{align}
where we've labeled the zero-modes as $h_u = H^{1(0)}_1, h_d = H^{2(0)}_2, \overline{\psi}_h = \overline{\psi}^{2(0)}_L, \psi_h = \psi^{1(0)}_R$.

As previously discussed, we may still choose where to define our matter fields. We may either keep them restricted to the $y=0$ brane or allow them to propagate in the 5D bulk. Restricting them to the brane results in the MSSM at low energies with supersymmetry breaking masses given by
\begin{align}
 m_{1/2} &= \hat{\alpha},
& \mu &= \hat{\gamma}, \nonumber \\
 m^2_{h_u, h_d} &= \hat{\alpha}^2,
& \mu B &= -2 \hat{\alpha} \hat \gamma,
\label{eq:soft_masses_SU5}
\end{align}
and
\begin{align}
m^2_{\tilde{q}, \tilde{u},\tilde{d}, \tilde{l}, \tilde{e}} &= 0,
&  A_0 &= - \hat{\alpha},
\label{eq:squark_masses_brane}
\end{align}
where we take the GUT scale $M_{GUT}$ as the compactification scale $M_{GUT} = 1/R$ and define $\hat \alpha = \alpha/R$ and $\hat \gamma = \gamma/R$. Note that with the brane matter placement the trilinear $A_0$ still gets a contribution from the $\partial_5 H^2 (d W / d H^1)$ term in Eq.~(\ref{SuperPotential}).

We stress that these constraints on the supersymmetry breaking parameters are particular to how the model is defined in the extra dimension. Indeed, if we instead allow matter in the bulk we gain extra contributions to $A_0$ and the squark soft SUSY breaking masses which arise from the $SU(2)_R$ symmetry. Then we have soft masses as seen in Eq.~(\ref{eq:soft_masses_SU5}), but now have,
\begin{align}
m^2_{\tilde{q}, \tilde{u},\tilde{d}, \tilde{l}, \tilde{e}} =& \hat \alpha^2,
&  A_0 =& - 3\hat{\alpha}.
\label{eq:squark_masses_bulk}
\end{align}
The extra contributions to $A_0$ and the soft SUSY breaking squark masses  arise as a consequence of the matter fields transforming under the $SU(2)_R$ symmetry.

With $1/R \sim 10^{16}\,$GeV, this model will naturally produce a supersymmetry breaking scale of order the GUT scale, far too high for low energy supersymmetry. In \cite{Barbieri1} the authors set $\alpha$ and $\gamma$ to be extremely small, so that $\hat \alpha = \alpha/R$ and $\hat \gamma = \gamma/R$ are of order a TeV. Consequently $\alpha$ and $\gamma$ must be of order $10^{-13}$, which presents a fine-tuning problem. Why must they be so small but non-zero? It seems that we have just swapped the gauge hierarchy problem for another fine-tuning problem. We will not tackle this issue here, but only express hope that these small parameters may be caused by the underlying UV completion of the theory.

 We will also restrict ourselves here to models with a reasonably low supersymmetry breaking scale so that the hierarchy problem itself is not an issue. Therefore we make the restriction that $\hat \alpha$ be less than $10^4$ GeV. We allow $\tan \beta$ to vary from 1 to 40. Once our low energy scenarios are generated in FlexibleSUSY we then confront them  with  the Scherk-Schwarz condition and the experimental constraints outlined in Sec.~\ref{sec:constraints},

\begin{figure}[htb!]
 \includegraphics[scale = 0.35]{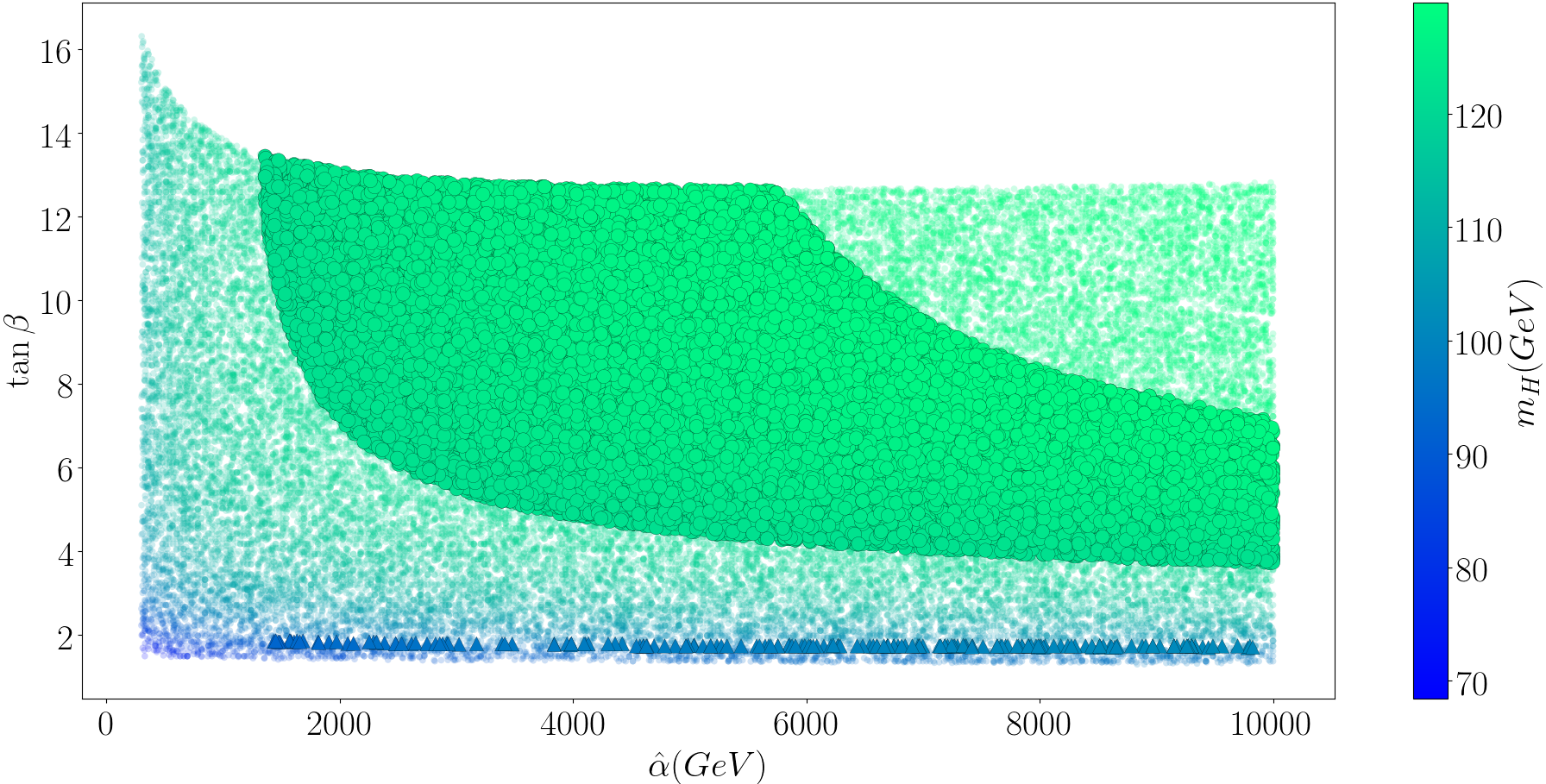}
 \caption{Points from the Barbieri, Hall and Nomura SU(5) Model with brane matter. Circles denote points that have passed the experimental constraints and have the desired Higgs mass;  triangles show points that obey the Scherk-Schwarz constraint. Fainter points in the background fail these constraints but are otherwise well behaved.}\label{SU5brane}
\end{figure}

\begin{figure}[htb!]
 \includegraphics[scale = 0.35]{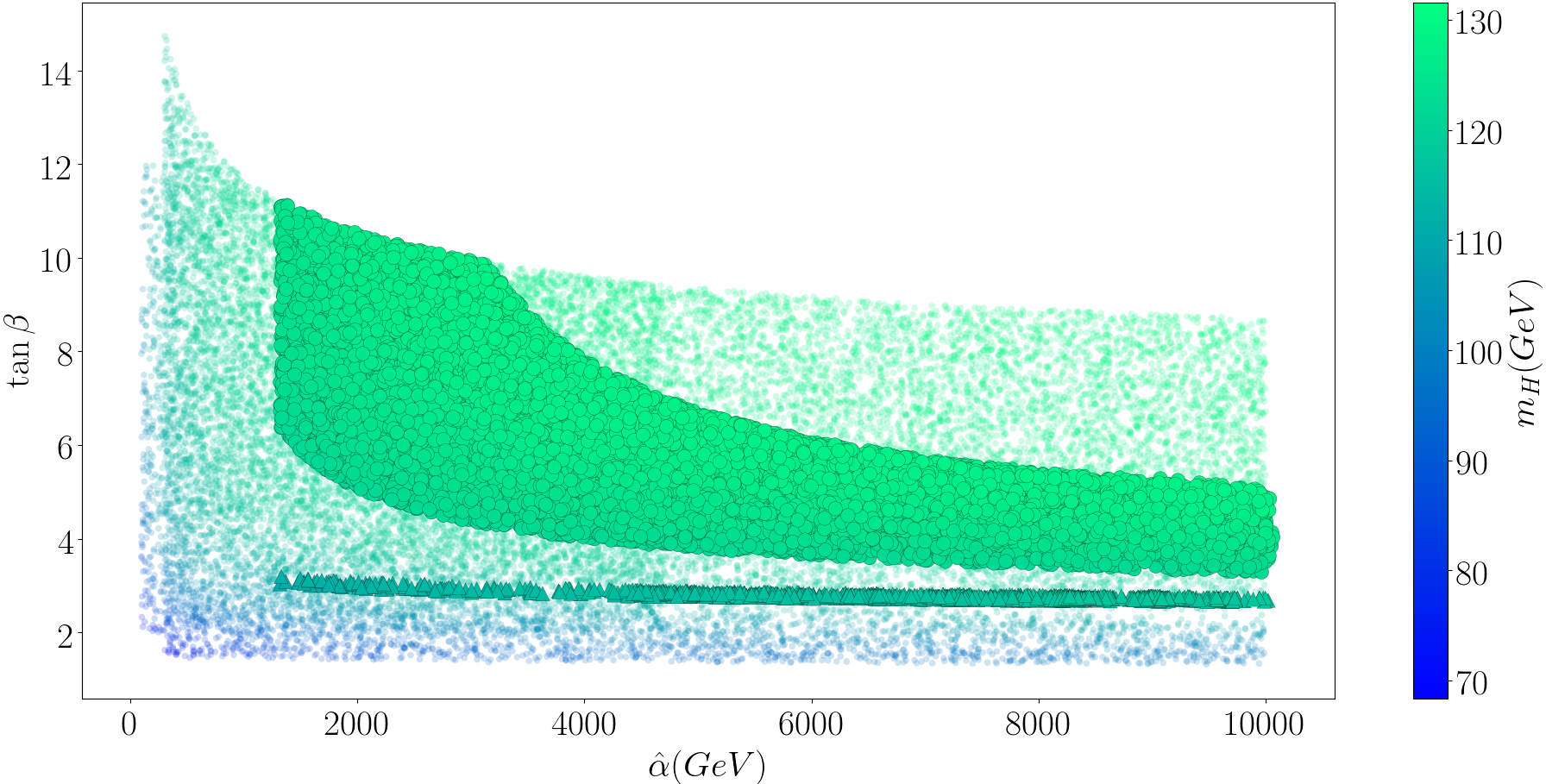}
 \caption{Points from the Barbieri, Hall and Nomura SU(5) Model with bulk matter. We use the same conventions for the points as in Figure~\ref{SU5brane}.}\label{SU5bulk}
\end{figure}

Since we have allowed $\tan \beta$ to fix $\hat \gamma$ our only input parameters are $\hat \alpha$ and $\tan \beta$. We show generated scenarios in the $\hat{\alpha}$-$\tan\beta$ plane in Figures~\ref{SU5brane} and \ref{SU5bulk}. The colour bar represents the mass of the lightest Higgs boson which we would like to identify with the discovered $125\,$GeV resonance. Points denoted with a circle have passed the LHC and DM constraints, and have the desired Higgs mass. In contrast, points that pass the Scherk-Schwarz constraint are denoted by triangles. The fainter points in the background are points that fail these constraints (but are otherwise well behaved). Figure~\ref{SU5brane} shows scenarios where the matter is kept on the $y=0$ brane, while Figure~\ref{SU5bulk} allows matter to propagate in the bulk.

We see that there is no overlap between the points providing the correct Higgs mass, while passing LHC and DM constraints, and those that conform with the Scherk-Schwarz condition. In essence, the Scherk-Schwarz condition prohibits a heavy enough lightest Higgs boson. However, we note that the Higgs boson mass is not too far from its measured value, particularly when matter is allowed to propagate in the bulk, which encourages us to study non-minimal extensions. We also note that the points with acceptable Higgs mass allow lower $\tan \beta$ as $\hat \alpha$ is increased, so there may still be room for agreement with the SS constraint in theories of High Scale or Split Supersymmetry which allow higher values of $\hat \alpha$ (see for example Ref.~\cite{Giudice:2011cg}).

It is difficult to provide a definitive explanation of why the Scherk-Schwarz condition requires such a low value of $\tan \beta$, since $\tan \beta$ is a low energy parameter arising from electroweak symmetry breaking involving parameters that are evolved from the high scale. Nevertheless we can obtain some understanding by examining the leading order tadpole equations, temporarily making the assumption that the supersymmetry breaking parameters don't evolve between the high and low scales. Then $m_{H_u}^2$, $m_{H_d}^2$, $\mu$ and $\mu B$ keep their values in terms of $\hat \alpha$ and $\hat \gamma$ seen in Eq.~(\ref{eq:soft_masses_SU5}). Plugging these into the leading-order tadpoles gives
\begin{gather}
  \hat{\alpha}^2 + \hat{\gamma}^2 - 2 \hat{\alpha} \hat{\gamma} \frac{1}{\tan \beta} - \frac{1}{8} \nu^2 (g_1^2 + g_2^2) \frac{1- \tan^2 \beta}{1 + \tan^2 \beta} = 0 ,\\
  \hat{\alpha}^2 + \hat{\gamma}^2 - 2 \hat{\alpha}\hat{\gamma} \tan \beta + \frac{1}{8} \nu^2 (g_1^2 + g_2^2) \frac{1- \tan^2 \beta}{1 + \tan^2 \beta} = 0.
\end{gather}
The sum of these has a solution $\tan \beta =\hat \alpha/\hat \gamma$, while the difference leads to $\hat \alpha = \hat \gamma$. Therefore at leading order and with no running, we would always expect $\tan \beta = 1$. Clearly the supersymmetry breaking parameters {\em do} run, and our tadpole equations are taken beyond leading order,
\[
  \frac{\partial V}{\partial \phi_i} - \frac{1}{\nu_i} \frac{\partial \Delta V}{\partial \phi_i} = 0,
\]
 where the corrections are stated in \cite{Dedes:2002dy}. So we will deviate from this expectation, but this provides some justification for why the Scherk-Schwarz boundary conditions may lead to a low value of $\tan \beta$. 

\section{An SU(5) model with an additional singlet} \label{sec:su5singlet}

We have seen that a minimal SU(5) model does not support Higgs bosons heavy enough to be the observed 125~GeV resonance. However, the Higgs boson mass may gain contributions from additional states in the spectrum, so we now extend our investigation by considering the model with an additional scalar electroweak singlet.

We have two choices for introducing the new scalar: we could introduce a chiral multiplet scalar singlet on the brane $S = (s, \psi_s)$; or introduce a hypermultiplet $\mathscr{S} = \{s^i, \Psi_S \}$ coupled to the Higgs. Here we will only couple our scalar to the Higgs and to itself, but again consider having matter in both the brane or the bulk.

The most general next-to-minimal superpotential that will result in either of the scalar/matter combinations at the low energy will be that of a general NMSSM:
\begin{equation}
 W = W_{\text{Higgs-Fermions}} + \mu H_u H_d + \lambda H_u H_d S + \frac{1}{3} \kappa S^3 + L S + \frac{1}{2} M_S S^2\label{WHiggs_5S}
\end{equation}

Note that we have kept an explicit $\mu H_u H_d$ term in contrast to the more usual $\mathds{Z}_3$-invariant NMSSM for which this term is absent. This is because the model does indeed produce an effective $\mu$ via the $\partial_5$ derivative, thus breaking the $\mathds{Z}_3$ symmetry of the NMSSM. Using a shift symmetry we set the linear term $L = 0$, and will also set $M_S = 0$, not to be confused with $m_S^2$ the soft SUSY breaking mass for the scalar superfield.

Our effective holomorphic terms are then a combination of the Scherk-Schwarz $SU(2)_H$ flavor breaking along with a contribution arising from the vev of $S$,
\begin{align}
 \mu_{\text{eff}}   & = \mu  + \frac{1}{\sqrt{2}} \lambda \expval{S}, \\
 \mu B_{\text{eff}} & = \mu B + \frac{1}{\sqrt{2}} T_{\lambda} \expval{S} + \frac{1}{2} \kappa \lambda \expval{S}^2.
\end{align}
We also assume that the only soft SUSY breaking masses arise from the Scherk-Schwarz mechanism. Therefore our only additional input parameters are $\kappa$ and $\lambda$.

For the simplest case, with the scalar $S$ on the brane as a chiral supermultiplet along with brane confined matter, we find soft SUSY breaking masses as in Eqs.(\ref{eq:soft_masses_SU5}) and (\ref{eq:squark_masses_brane}), and additionally
\begin{equation}
 m_S^2 = 0	\qquad T_{\lambda} = -2\lambda\hat{\alpha}	\qquad	T_{\kappa} = 0.
 \label{eq:Smasses_brane}
\end{equation}
This new equation (\ref{eq:Smasses_brane}) holds also for a bulk fermions too but must be used with Eqs.(\ref{eq:soft_masses_SU5}) and (\ref{eq:squark_masses_bulk}).

\begin{figure}[thb!]
\includegraphics[scale = 0.35]{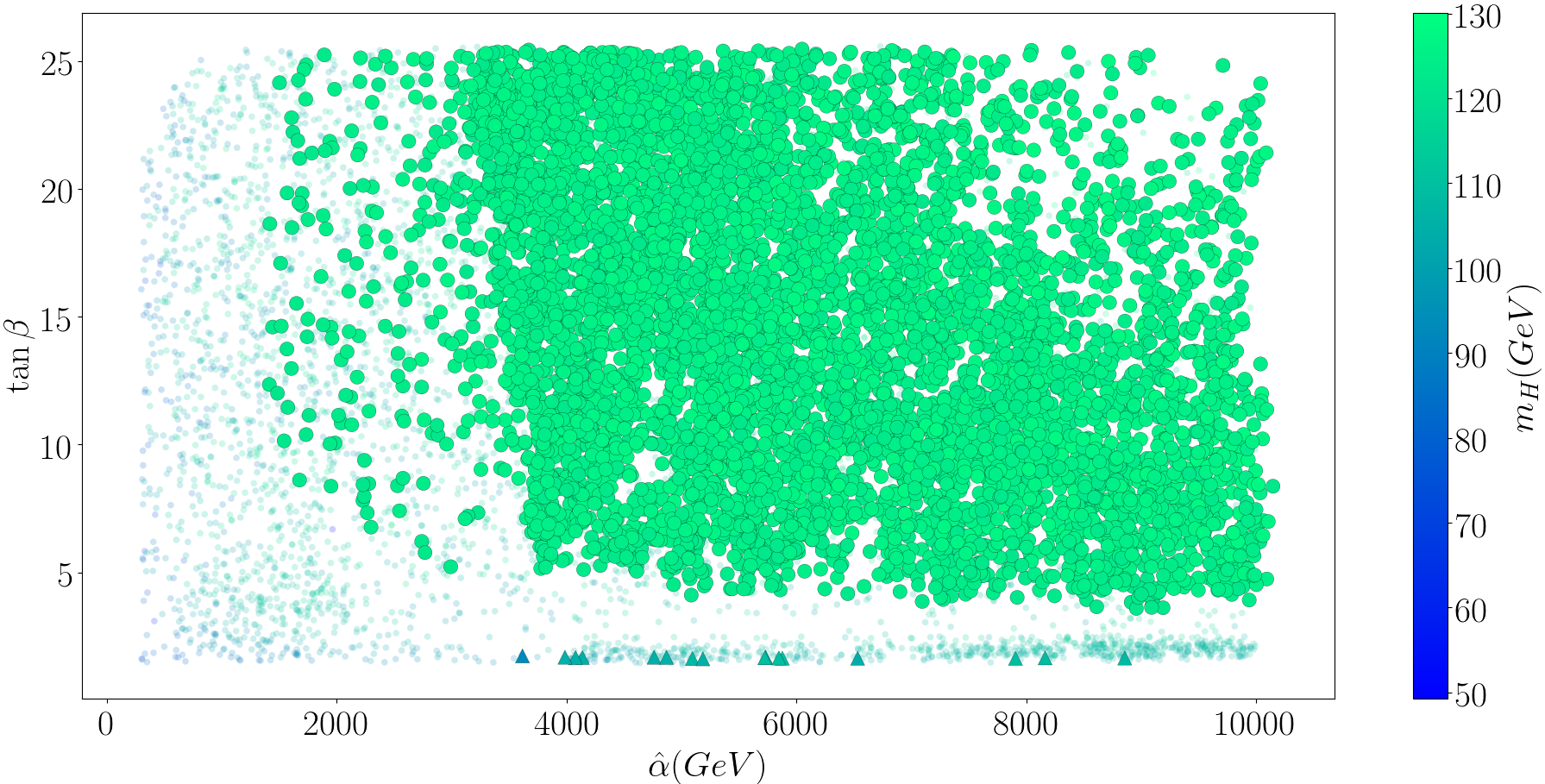}
 \caption{Points for the $SU(5)$ model with an additional scalar $S$ on the brane, and brane matter. We use the same conventions for the points as in Figure~\ref{SU5brane}.}\label{SMSSM_BrS_BrMat}
\end{figure}

\begin{figure}[thb!]
 \includegraphics[scale = 0.35]{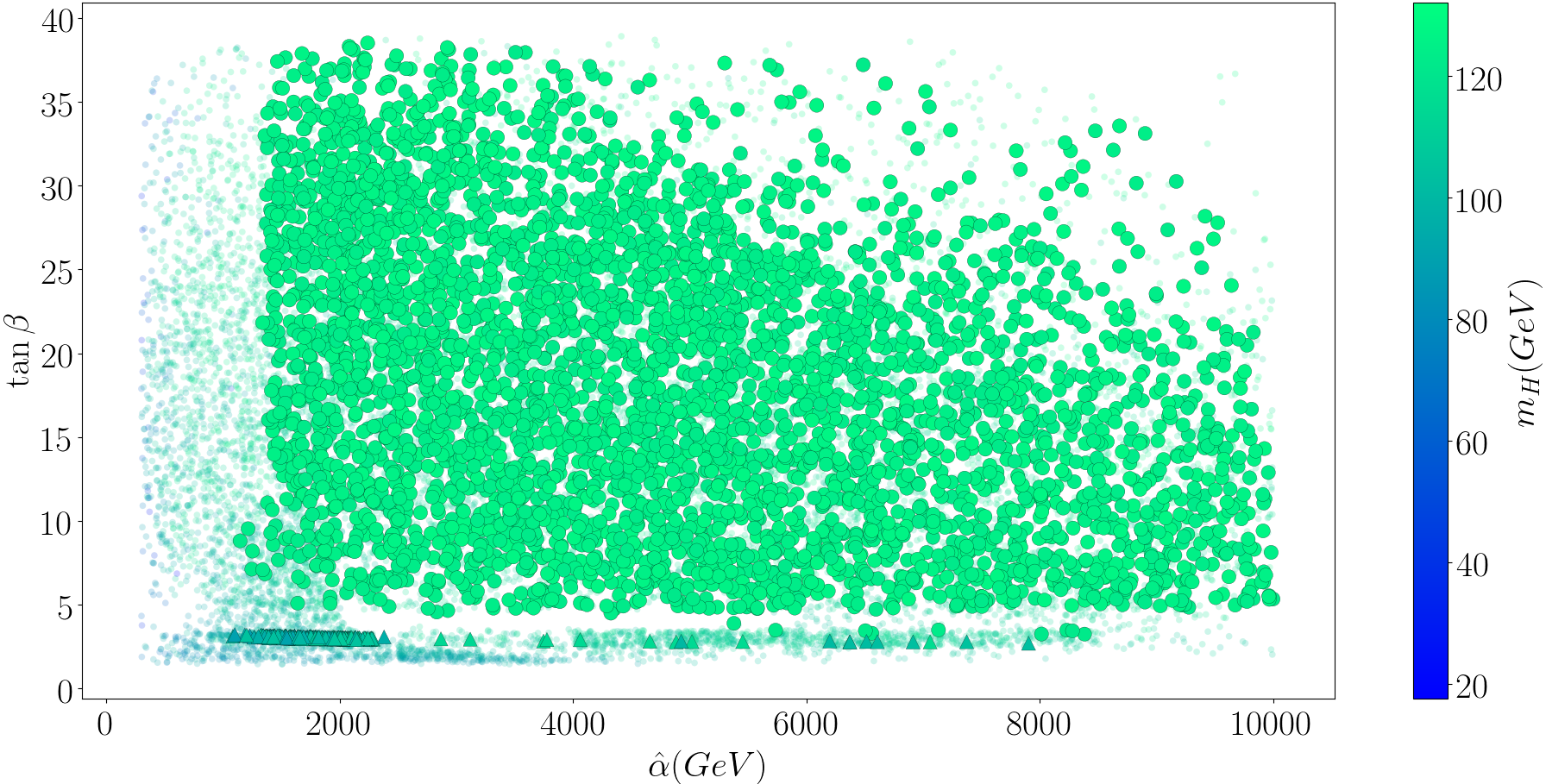}
 \caption{Points for the $SU(5)$ model with an additional scalar $S$ on the brane, and bulk matter. We use the same conventions for the points as in Figure~\ref{SU5brane}.}\label{SMSSM_BrS_BuMat}
\end{figure}

The results or our analysis for this model are shown in Figures \ref{SMSSM_BrS_BrMat} and \ref{SMSSM_BrS_BuMat}, where we using the same convention for the points as in the previous figures. We allowed the additional parameters $\lambda$ and $\kappa$ to vary from $0$ to $0.9$. We see that without enforcing the Scherk-Schwarz constraint, both versions produce an appropriate low energy SM spectrum with the appropriate Higgs mass. The only significant difference is that bulk matter allows a larger range of $\tan\beta$ values, while the brane matter requires $\tan \beta \lesssim 25$. It is also interesting to note that all the acceptable points reside in the region with  $\hat{\alpha} \gtrsim 2\,$TeV indicating that these correspond to a SUSY scale that `naturally' falls in the $\mathcal{O}(1-10)$\SI{}{TeV} range.

However, as for the `vanilla' $SU(5)$ model, the points that pass the Scherk-Schwarz constraint do not overlap with those which pass Higgs and LHC constraints. The contribution to the Higgs mass from the additional singlet has not been sufficient to provide agreement. This is a recurrent theme that we see all through our studies; the points that would originate from the Scherk-Schwarz breaking of $SU(2)_H$, $SU(2)_R$, have difficulty producing a large enough Higgs mass and/or don't pass LHC constraints. This is more pronounced when we have fermions on the brane than when they are in the bulk, but remains true in both cases. Indeed, in the latter case, the Schrek-Schwarz constraint comes rather close to the acceptable phenomenological region, so a closer look is needed. Perhaps we have been overly conservative with our error estimates for $\mu, \mu B$, and a relaxation of these uncertainties would allow agreement.
For example, the maximum Higgs mass for the Scherk-Schwarz points is $m_H \approx \SI{116.9}{GeV}$, which is close enough to provide some doubt.

Alternatively we may place the additional scalar in the bulk. To achieve this, we introduce the $SU(5)$ singlet hypermultiplet $\mathscr{S} = \{ s^i, \Psi_s \}, \Psi_s = (\psi^s_L, \psi^s_R)$, where $s^i, i=1, 2$ transforms only under the $SU(2)_R$ residual supersymmetry. Analogous to our previous treatment, we assign the $Z$ parities,
\begin{align}
 Z = +1 & : 			\qquad	s^1, \psi^s_L, \\
 Z = -1 & :			\qquad	s^2, \psi^s_R.
\end{align}
This projects out the corresponding zero-modes, which are then coupled to our Higgs in the same way as in Eq.~(\ref{WHiggs_5S}). Under $T$, the fields transform according to,
\begin{equation}
 \begin{pmatrix}	s^1 \\ s^2	\end{pmatrix} = e^{i \alpha \sigma^2 y /R}  \begin{pmatrix}	\tilde{s}^1 \\ \tilde{s}^2	\end{pmatrix},
\end{equation}
which again will produce a soft SUSY breaking mass for the scalar $m_S^2$, via the $\partial_5$.

In this case of a bulk scalar hypermultiplet $\mathscr{S}$ we again have Eq.~(\ref{eq:soft_masses_SU5}), and either Eq.~(\ref{eq:squark_masses_brane}) or Eq.~(\ref{eq:squark_masses_bulk}) for either brane or bulk fermions respectively. However, instead of Eq.~(\ref{eq:Smasses_brane}) we now have
\begin{equation}
 m_S^2 = \hat{\alpha}^2	\qquad T_{\lambda} = -3\lambda\hat{\alpha}	\qquad	T_{\kappa} = -\kappa \hat{\alpha}. \label{eq:Smasses_bulk}
\end{equation}

\begin{figure}[htb!]
 \includegraphics[scale = 0.35]{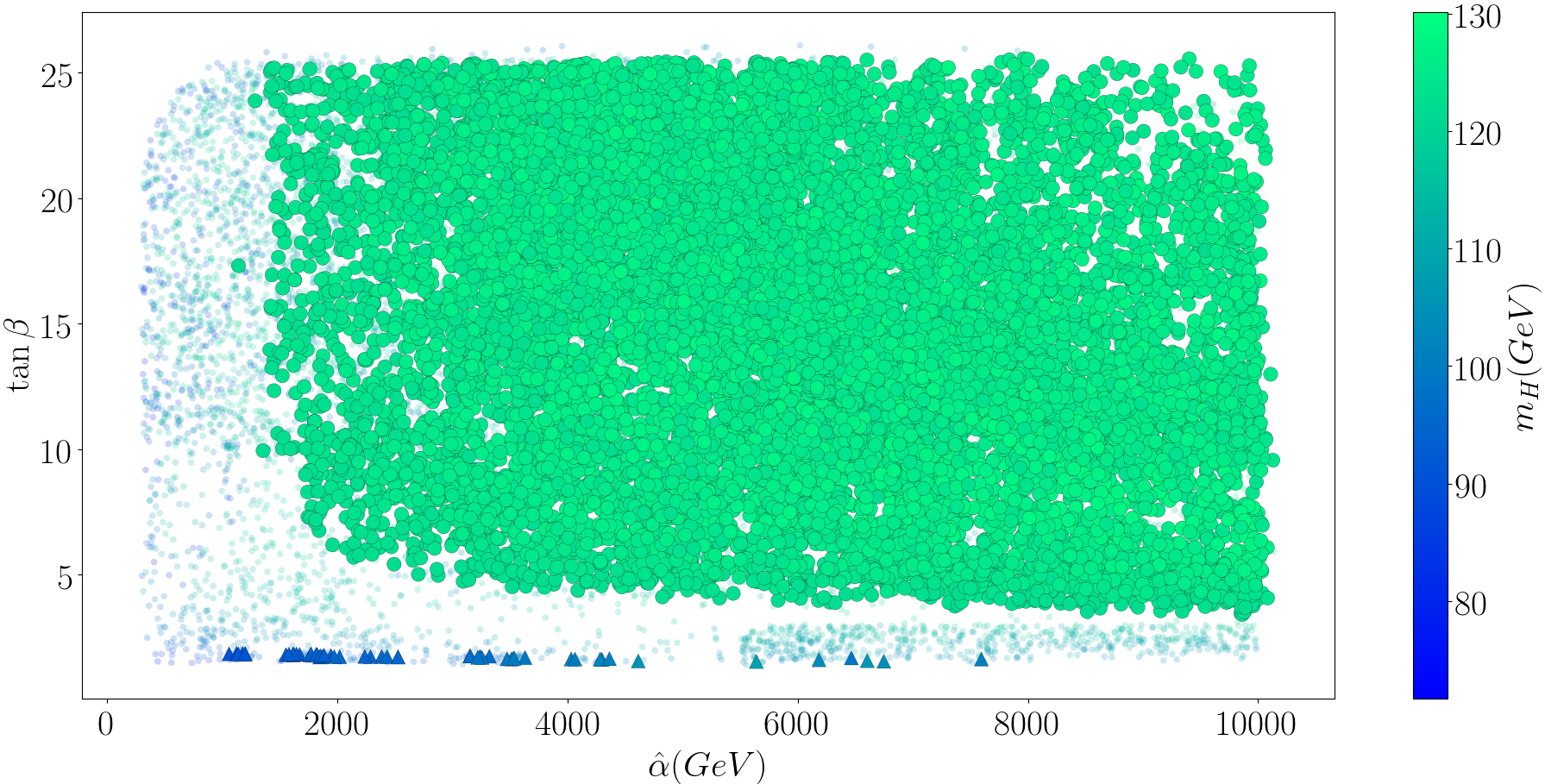}
 \caption{Points for the $SU(5)$ model with an additional bulk scalar $\mathscr{S}$, and brane matter. We use the same conventions for the points as in Figure~\ref{SU5brane}.}\label{SMSSM_BuS_BrMat}
\end{figure}

\begin{figure}[htb!]
 \includegraphics[scale = 0.35]{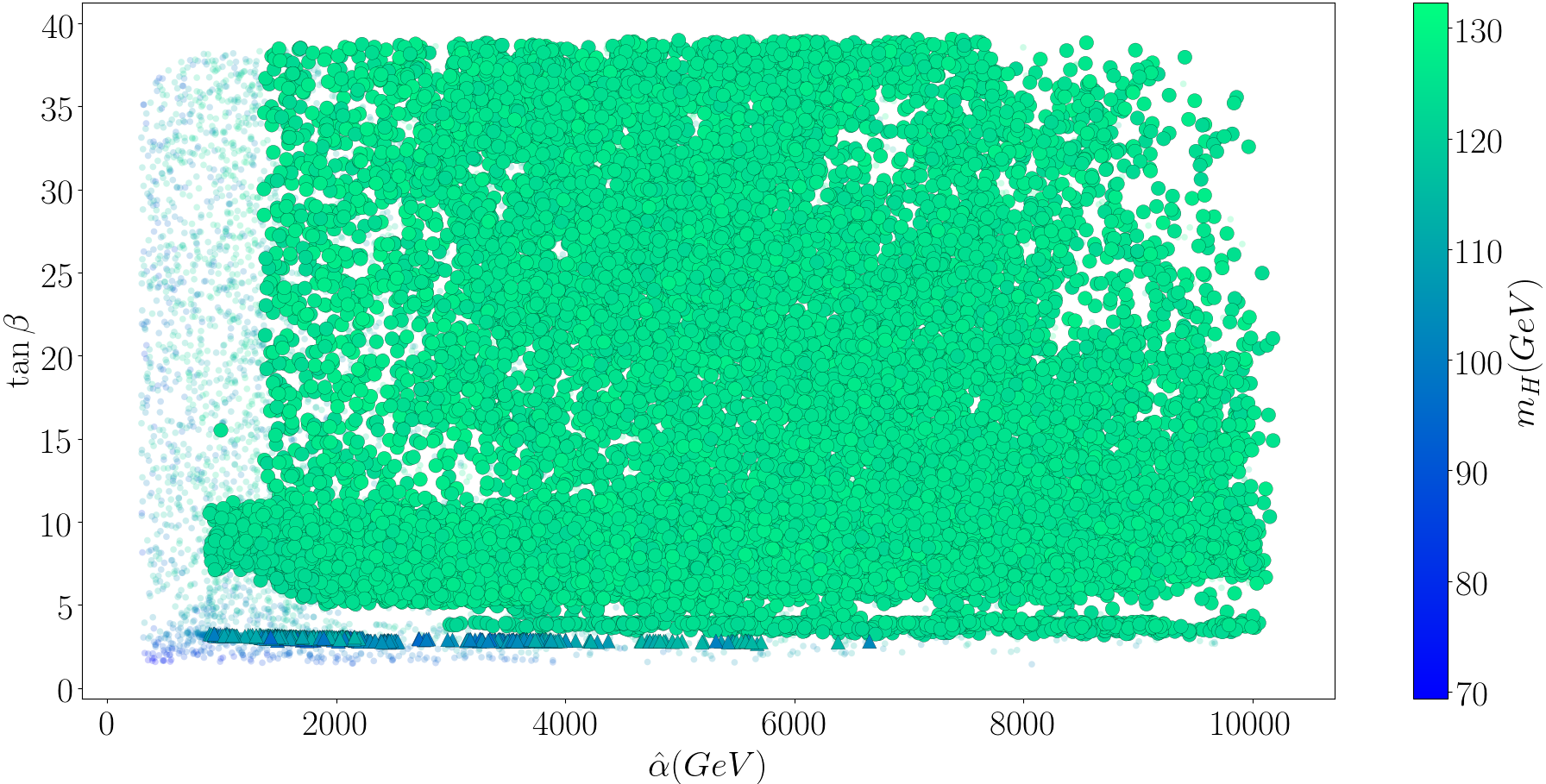}
 \caption{Points for the $SU(5)$ model with an additional bulk scalar $\mathscr{S}$, and bulk matter. We use the same conventions for the points as in Figure~\ref{SU5brane}.}\label{SMSSM_BuS_BuMat}
\end{figure}

\begin{figure}[thb!]
 \includegraphics[scale = 0.4]{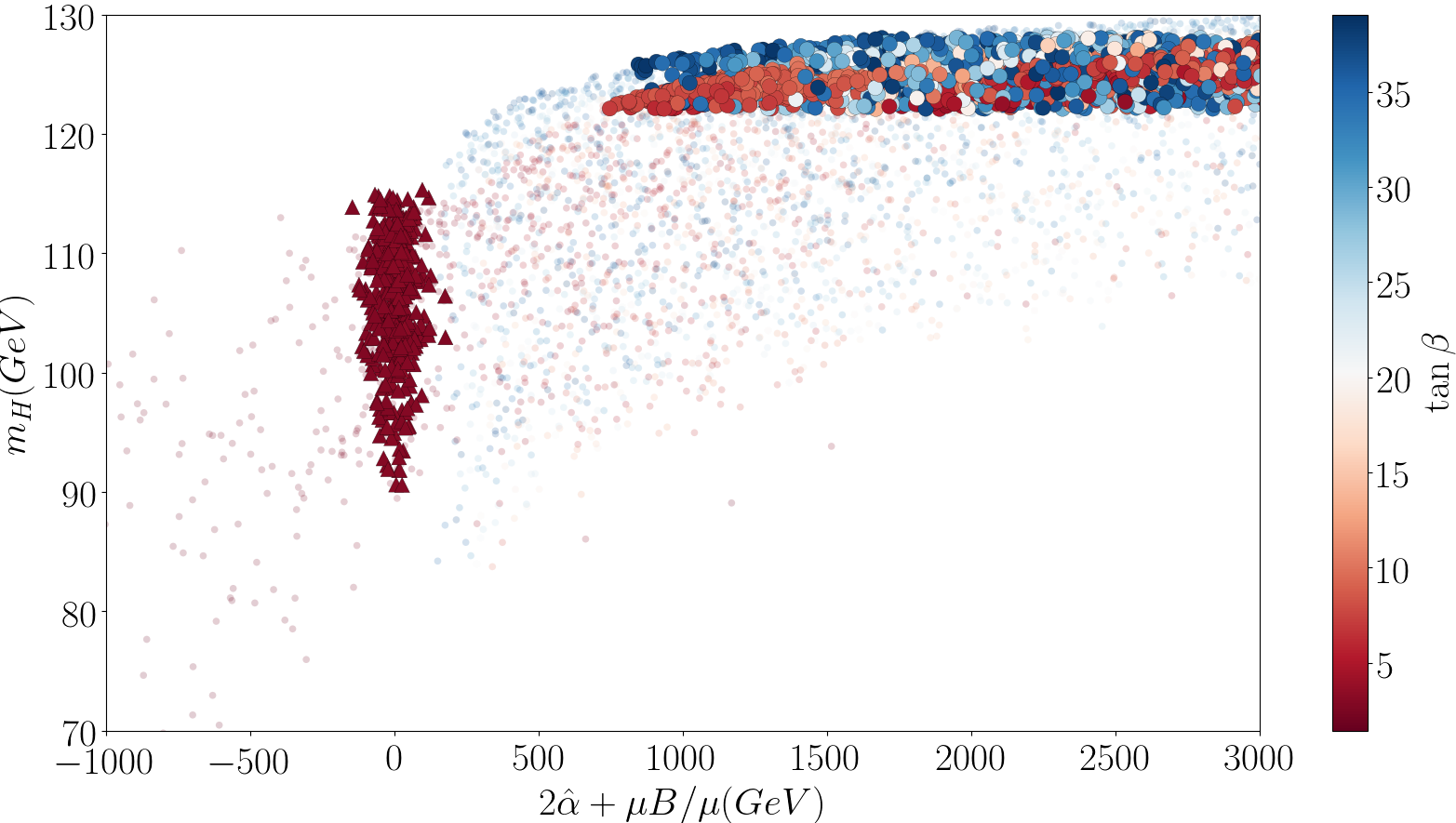}
 \caption{Points for the $SU(5)$ model with an additional bulk scalar $\mathscr{S}$, and bulk matter. We use the same conventions for the points as in Figure~\ref{SU5brane}, but this time we have plotted the deviation from the Scherk-Schwarz condition vs.\ the Higgs mass $m_H$.}\label{SMSSM_BuS_BuMat_Alt1}
\end{figure}

The results for these choices are shown in Figures \ref{SMSSM_BuS_BrMat} and \ref{SMSSM_BuS_BuMat}. Our story seems to repeat itself as the Scherk-Schwarz condition is not compatible with the Higgs mass and/or LHC constraints. Once again, the gap is much more pronounced for brane matter than bulk matter, and indeed the gap looks almost absent in Figure \ref{SMSSM_BuS_BuMat}. To be clear that there is indeed no overlap in this latter case, we have also plotted the data of Figure \ref{SMSSM_BuS_BuMat} as $2 \hat \alpha + \mu B/\mu$ against $m_H$, with $\tan \beta$ as the point's colour in Figure \ref{SMSSM_BuS_BuMat_Alt1}. The Schrek-Schwarz condition is exactly realised for points at $2 \hat \alpha + \mu B/\mu=0$ and the spread of points around this value is due to uncertainties. One can clearly see that these points have no overlap with the correct Higgs mass region. Even when we artificially inflate our uncertainties by a factor of 10 (not shown), we do not find an overlap, though the Higgs mass becomes significantly better. So unfortunately once again we cannot reconcile this model and Scherk-Schwarz breaking with the Higgs mass and experimental constraints.

We may also consider a variant of this model similar to the more usual $\mathds{Z}_3$-invariant NMSSM but setting our $\mu=0$, so that the superpotential is
\begin{equation}
 W = W_{\text{Higgs-Fermions}}(\mu = 0) + \lambda H_u H_d S + \frac{1}{3} \kappa S^3 + L S + \frac{1}{2} M_S S^. \label{WHiggs}
\end{equation}

Here we have effectively set $\hat \gamma=0$ and allowed an effective $\mu$ and its supersymmetry breaking partner parameter to be generated entirely throught the vev of the new scalar. That is,
\begin{align}
 \mu_{\text{eff}}    & =  \frac{1}{\sqrt{2}} \lambda \expval{S}, \\[2mm]
 \mu B _{\text{eff}} & =  \frac{1}{\sqrt{2}} T_{\lambda} \expval{S} + \frac{1}{2} \kappa \lambda \expval{S}^2.
\end{align}
Significantly, since $\hat \gamma=0$, the Scherk-Schwarz constraint is absent and electroweak symmetry breaking proceeds just like in the NMSSM with freedom to choose $\tan \beta$. However, this has extremely constrained supersymmetry breaking parameters since $\hat{\alpha}$ is the only input at high energies.

Unfortunately, we now find that we are unable to simultaneously satisfy this restrictive high scale boundary condition and the electroweak tadpole constraints, irrespective of our choice of brane/bulk scalars or brane/bulk fermions. The model is simply too constrained and is not viable. One could imagine introducing an additional scalar mass $M_S$ and associated $B_{M_S}$ to increase the freedom of the model, possibly allowing the correct pattern of electroweak symmetry breaking, but this is beyond the philosophy of our study because it introduces additional supersymmetry breaking by hand, separate from the SS mechanism.

\section{An SU(5) $\times$ U(1) model} \label{sec:su5xu1}

Confronted with the inability of the simplest models to generate a heavy enough Higgs boson while avoiding LHC and Dark Matter constraints, we may once again increase the complexity of our model. Next we will consider an $SU(5)$ model with an additional $U(1)$ symmetry, similar to the USSM. The superpotential is identical to that of Eq.~(\ref{WHiggs}), but with the added complexity of the low energy gauge group being extended to $G_{SM} \times U(1)$. The additional $U(1)$ is broken at the SUSY scale via the brane scalar (projected or placed), prior to which we assign our fields appropriate charges. The assignment of these charges is arbitrary and model dependent, but it is useful to set them to correspond with those arising from embedding in some larger group such as $SO(10)$ or $E_6$ (see section~\ref{sec:e6} for more details).

As an example, we will choose the $E_6$ inspired charge assignments. The $5D$ bulk gauge structure is assumed to be $SU(5) \times U(1)$, and as in the previous examples, the Schrek-Schwarz compactification, with $Z$ and $T$ unchanged, will break this gauge group on the brane down to $G_{SM} \times U(1)$. The $E_6$ inspired charges under the $U(1)$ group are \cite{King:2005my},
\begin{equation}
\begin{array}{rlrlrlrlrl}
 Q_q \!\!\! &= \displaystyle \frac{1}{\sqrt{40}}, \quad &
 Q_l \!\!\! &= \displaystyle \frac{2}{\sqrt{40}},	\quad &
 Q_d \!\!\! &= \displaystyle \frac{2}{\sqrt{40}}, \quad &
 Q_u \!\!\! &= \displaystyle \frac{1}{\sqrt{40}}, \quad	&
 Q_e \!\!\! &= \displaystyle \frac{1}{\sqrt{40}},	\\[4mm]
&&
 Q_{H_d} \!\!\! &= \displaystyle - \frac{3}{\sqrt{40}},	\quad &
 Q_{H_u} \!\!\! &= \displaystyle -\frac{2}{\sqrt{40}}, \quad	&
 Q_S \!\!\! &= \displaystyle \frac{5}{\sqrt{40}}.
\end{array}
\end{equation}

The high scale boundary conditions and soft SUSY breaking masses remain as those for the $SU(5)$ model with an additional scalar in section~\ref{sec:su5singlet}, namely Eq.~(\ref{eq:soft_masses_SU5}) and the appropriate choice of Eqs.~(\ref{eq:squark_masses_brane}), (\ref{eq:squark_masses_bulk}), (\ref{eq:Smasses_brane}) and (\ref{eq:Smasses_bulk}), depending on the choice of whether the scalar and the fermions are placed on the brane or in the bulk. The difference in the spectra arises due to the presence of the extra $U(1)_N$ which will modify the RGEs. In addition the breaking of $U(1)_N$ will produce a $Z'$ boson, and we exclude points that violate the ATLAS bounds \cite{ATLASEXOTICS}.

\begin{figure}[htb!]
 \includegraphics[scale = 0.35]{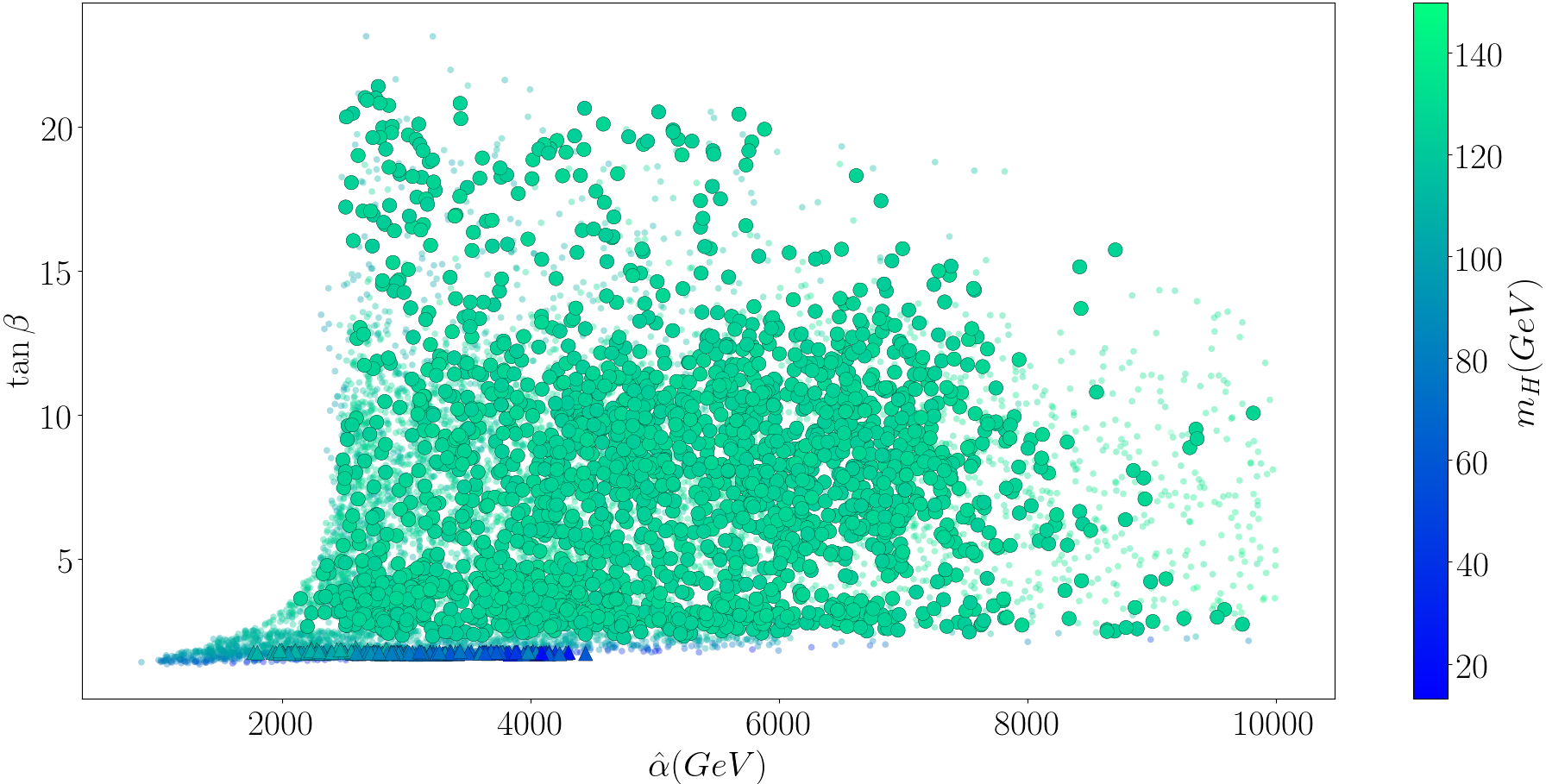}
 \caption{Points for the $SU(5)\times U(1)$ model with the additional scalar $S$ and fermions both on the brane. We use the same conventions for the points as in Figure~\ref{SU5brane}.}\label{UMSSM_BrS_BrMat}
\end{figure}

\begin{figure}[htb!]
 \includegraphics[scale = 0.35]{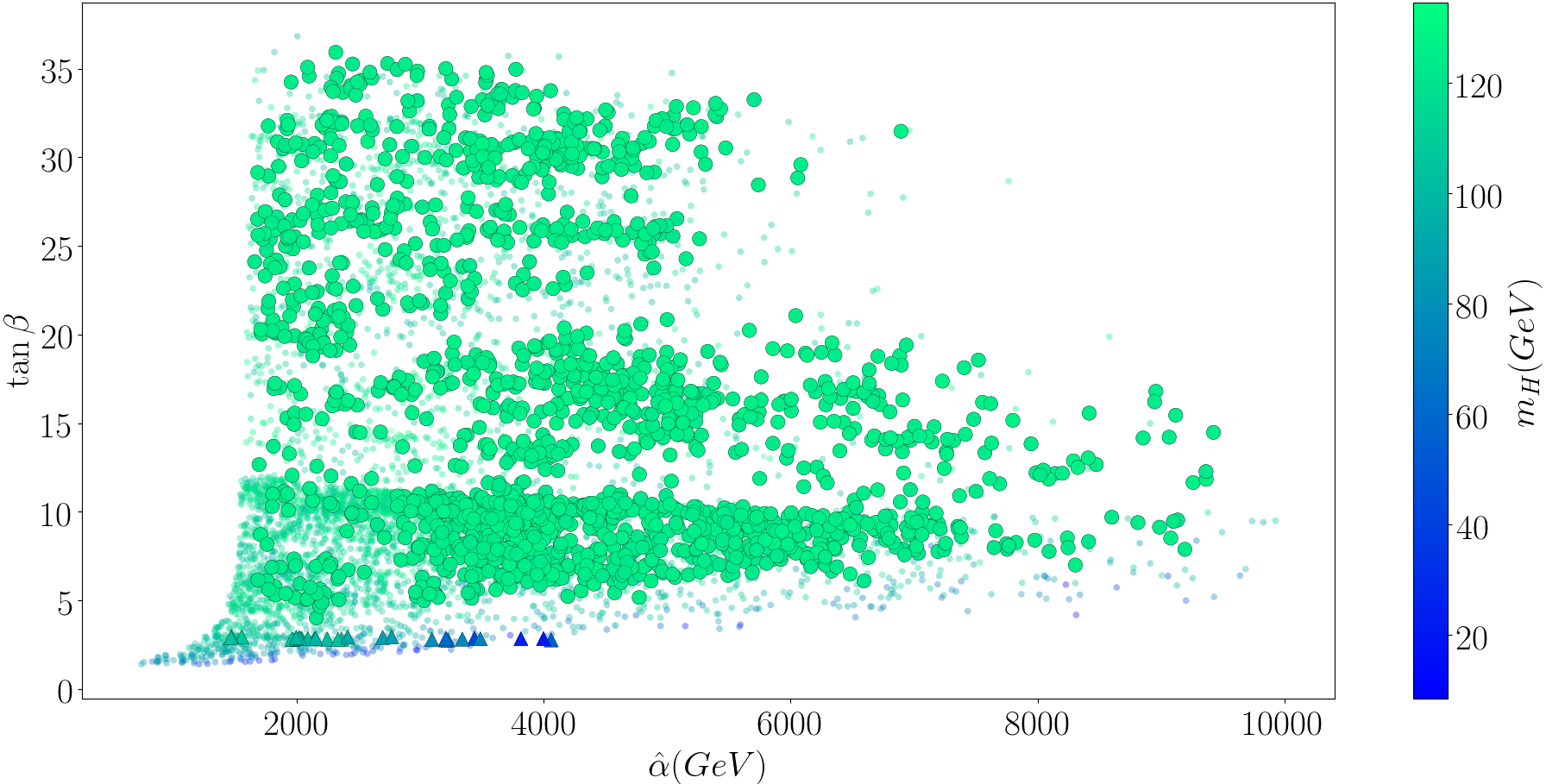}
 \caption{Points for the $SU(5)\times U(1)$ model with the additional scalar $S$ on the brane and fermions in the bulk. We use the same conventions for the points as in Figure~\ref{SU5brane}.}\label{UMSSM_BrS_BuMat}
\end{figure}

\begin{figure}[htb!]
 \includegraphics[scale = 0.35]{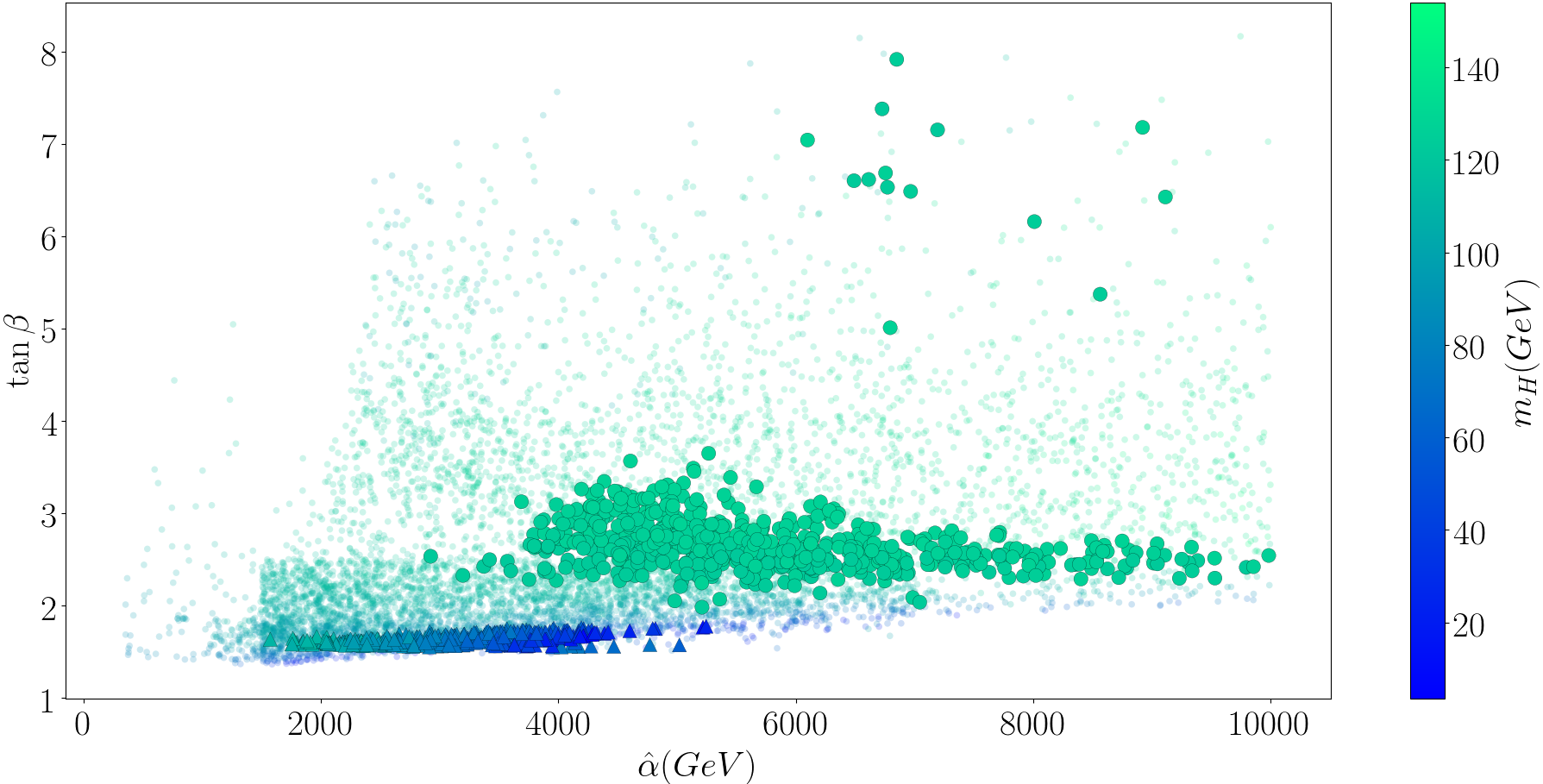}
 \caption{Points for the $SU(5)\times U(1)_N$ model with the additional scalar $\mathscr{S}$ in the bulk and fermions on the brane. We use the same conventions for the points as in Figure~\ref{SU5brane}.}\label{UMSSM_BuS_BrMat}
\end{figure}

\begin{figure}[htb!]
 \includegraphics[scale = 0.35]{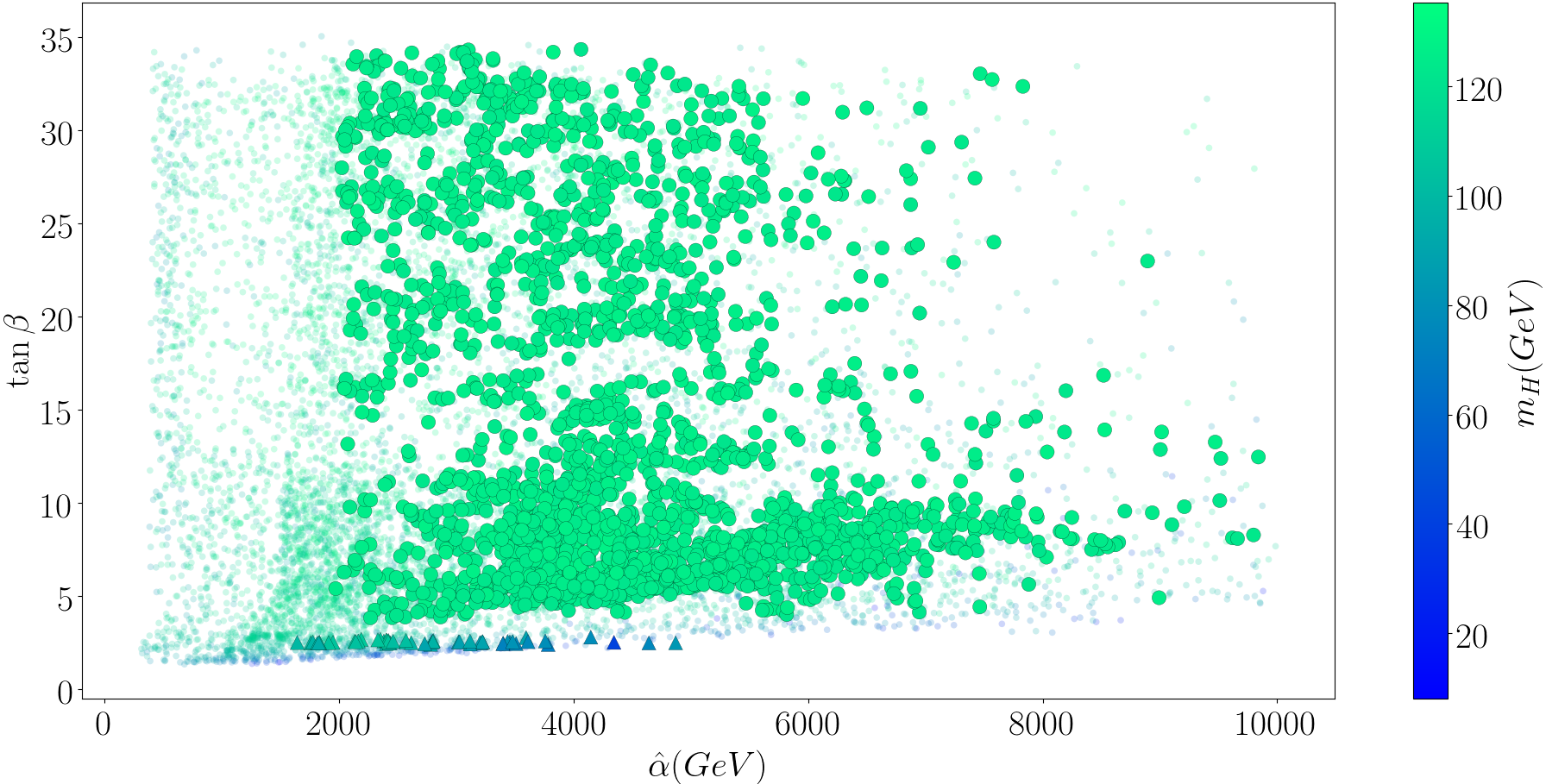}
 \caption{Points for the $SU(5)\times U(1)$ model with the additional scalar $\mathscr{S}$ and fermions both in the bulk. We use the same conventions for the points as in Figure~\ref{SU5brane}.}\label{UMSSM_BuS_BuMat}
\end{figure}

Performing our parameter scans for the additional scalar on the brane gives Figures~\ref{UMSSM_BrS_BrMat} and \ref{UMSSM_BrS_BuMat}, for brane or bulk fermions respectively. We see many regions that pass LHC and Dark Matter constraints, but again the points passing the Scherk-Schwarz constraint do not overlap, though they come close when the scalar is on the brane and fermions are in the bulk case, only finally being excluded by the LHC constraints. This pattern repeats if we allow the additional scalar into the bulk, as seen in Figures \ref{UMSSM_BuS_BrMat} and \ref{UMSSM_BuS_BuMat}. It is interesting that with the additional scalar $S$ in the bulk and fermions on the brane, the constraints favour scenarios with lower $\tan \beta$.

We also examined the same model with $\mu$ explicitly set to $0$ (so $\hat \gamma=0$), as we did for the model in section~\ref{sec:su5singlet}, to bypass the Scherk-Schwarz constraint. Unfortunately no placement of our fields on brane or bulk were able to produce scenarios with EWSB.

Of course, the setting of our $U(1)$ charges need not follow the pattern of $E_6$, as the $U(1)$ may be of some completely different origin. Another obvious example would be a $U(1)$ as a remnant of $SO(10)$, in which case the charge assignments would be \cite{Asaka},
\begin{equation}
\begin{array}{rlrlrlrlrl}
 Q_q \!\!\! &= -1, \quad &
 Q_l \!\!\! &= 3,	\quad &
 Q_d \!\!\! &= 1, \quad &
 Q_u \!\!\! &= 3, \quad	&
 Q_e \!\!\! &= -5,	\\[2mm]
&&
 Q_{H_d} \!\!\! &= -2,	\quad &
 Q_{H_u} \!\!\! &= 2, \quad	&
 Q_S \!\!\! &= 10.
\end{array}
\end{equation}
However, none of our models with these charge assignments, including fields in the bulk or on the brain and with $\mu$ set to $0$ or not, were able to provide satisfactory electroweak scale spectra.

\section{An $E_6$ model} \label{sec:e6}

The first model discussed in section~\ref{sec:su5xu1} carried the $U(1)$ charge assignments that might arise from a larger $E_6$ symmetry group. However, if the unification group were indeed $E_6$ one would expect other additional fields that may survive down to the electroweak scale. An example of such a model is the E$_6$SSM \cite{King:2005my,King:2005jy,Athron:2009bs,Athron:2012sq,Nevzorov}, which has a superpotential,
\begin{align}
 W_{E_6SSM} = & W_{MSSM}(\mu = 0) + \lambda H_u H_d S + \lambda_{\alpha\beta} S (H^d_\alpha) (H^u_\beta) + \kappa_{ij} S (D_{i} \overline{D}_j) + \tilde{f}_{\alpha\beta} S_\alpha (H^d_\beta H_u)  \nonumber \\
             & + f_{\alpha\beta} S_\alpha (H_d H^u_\beta) + g^D_{ij} (Q_i L_4) \overline{D}_j + h^E_{i\alpha} e^c_i(H^d_\alpha) + \mu_L L_4 \overline{L}_4
\end{align}
where $\alpha, \beta = 1, 2, 3$ and $i, j = 1, 2$ are generation indices. (For the definitions of these additional fields, see Ref.~\cite{Nevzorov}). Applying the Schrek-Schwarz high scale boundary conditions, with the $\mathbf{27}$ and $\overline{\mathbf{27}}$ representations placed in the bulk, gives,
\begin{equation}
 m_{1/2} = \hat{\alpha},
  \qquad m^2_{h_u, h_d,S,H^u_\alpha, H^d_\beta, D_i, \overline{D}_j, S_\alpha, L_4, \overline{L}_4} = \hat{\alpha}^2,
  \qquad T_{\xi} = -3 \xi \hat{\alpha}
\end{equation}
and Eq.~(\ref{eq:squark_masses_bulk}), where $\xi \in  \left\{\lambda, \kappa_{ij}, \lambda_{\alpha\beta}, \tilde{f}_{\alpha\beta}, f_{\alpha\beta}, g^D_{ij}, h^E_{i\alpha} \right\}$. In practice, we allow $\mu_L$ to vary independently, and set the values of $m^2_{H_d}$, $m^2_{H_u}$, $m^2_{S}$ during EWSB. We would then have to check for a new Scherk-Schwarz condition to make sure the full boundary conditions are obeyed. Unfortunately, even without enforcing this new Scherk-Schwarz condition, we find that the boundary conditions on the other parameters at the high scale are so restrictive that we can find no valid low energy scenarios.

We note that the implementation of this model is somewhat different from those described earlier because the the Higgs bosons themselves are in the $\mathbf{27}$ and $\overline{\mathbf{27}}$. Therefore the $SU(2)_H$ symmetry should be enlarged to encompass the full $\mathbf{27}$ and $\overline{\mathbf{27}}$. However, here we have taken the simplest route, ignoring this enlarged symmetry and allowing the holomorphic $\mu_{L} L_4 \overline{L}_4$ term to arise from some another unknown mechanism altogether (that is, allowing it to vary). It is possible that a more non-minimal implementation, where the $\mathbf{27}$ and $\overline{\mathbf{27}}$ symmetry is fully incorparated would have more luck in producing a viable phenomenology, but this is beyond the scope of this paper.

\section{Conclusions} \label{sec:conclusions}

In this investigation we have examined models of Scherk-Schwarz orbifold compactification. In these scenarios, the extra dimension of a $5D$ space is given periodic boundary conditions and rolled-up to a radius $R \sim 1/M_{\rm GUT}$; the space is folded to provide an orbifold with fixed points in the standard fashion. Scherk-Schwarz compactification differs from standard orbifold compactification in that it allows non-trivial transformations of the fields under the orbifolding symmetries. This Scherk-Schwarz orbifolding allows for the breaking of both supersymmetry and the GUT symmetry.

We apply this compactification to several models of Grand Unification, including $SU(5)$ unification, $SU(5)$ with an additional singlet, $SU(5) \times U(1)$, and an $E_6$ inspired model, all with several variations. The Scherk-Schwarz mechanism provides severe constraints on the supersymmetry breaking parameters at the unification scale. We apply these constraints and use Renormalisation Group equations to evolve the theory down to the electroweak scale, where they are confronted with low energy constraints from the LHC, the Dark Matter relic density and the Higgs mass.

We find that these boundary conditions are very difficult to combine with a $125\,$GeV Higgs boson. Generally, these models prefer a lighter Higgs boson and rather low $\tan \beta$, and despite and exhaustive scan and variations in the models we were unable to find parameter choices which simultaneously obeyed all low scale measurement constraints. In cases where the Higgs mass was in the correct range, for example in the $SU(5)$ models with an extra singlet when an effective Higgs-higgsino mass term was entirely generated by the Scherk-Schwarz breaking, the models were ruled out by other low energy constraints such as LHC chargino exclusions.

Although we studied several models with lots of variations, this work does not claim to rule out the Scherk-Schwarz compactification in general. One could imagine having more complicated gauge groups and extra-dimensional geometry which would change the unification constraints on the supersymmetry breaking masses. Indeed, we saw in the implementation of the E$_6$ gauge group that one additional freedom in allowing an alternative treatment of the large representations that now include the Higgs. However, we are confident in making the claim that Scherk-Schwarz compactification of $SU(5)$ models, $SU(5)$ models with an extra singlet, and $SU(5)\times U(1)$ models where the extra dimension is compactified on $S^1/\mathds{Z}_2$ are not compatible with electroweak scale obervations.

\section*{Acknowledgements}
DDS would like to thank Brian Alden for help with parallelising codes. DJM acknowledges partial support from the STFC grant ST/P000746/1.

\section*{References}
\bibliography{SSOrbifolds}
\bibliographystyle{JHEP}

\end{document}